\documentclass[letterpaper,twocolumn,10pt]{article}
\usepackage{usenix}

\usepackage{tikz}
\usepackage{amsmath}

\usepackage{amssymb}
\usepackage{booktabs}
\usepackage{multirow}
\usepackage{subcaption}
\usepackage{makecell}
\usepackage{graphicx} 
\usepackage{lipsum}
\usepackage{kotex}
\usepackage{color}
\usepackage{xcolor}
\usepackage{mdframed}
\usepackage{seqsplit}
\usepackage{xurl}
\usepackage{threeparttable}
\usetikzlibrary{arrows.meta, positioning, calc, fit}
\usepackage[most]{tcolorbox}

\usepackage{nicematrix}

\definecolor{cO0}{HTML}{9E9E9E}
\definecolor{cO1}{HTML}{4C9BE8}
\definecolor{cO2}{HTML}{E08E3C}
\definecolor{cO3}{HTML}{C0384F}

\newcommand{\grouplabel}[2]{%
  \rotatebox[origin=c]{90}{%
    \scriptsize\bfseries
    \shortstack{#1\\[-0.2ex]#2}%
  }%
}

\newcommand{\costbox}[1]{%
  \textcolor{#1}{\rule{1.5ex}{1.5ex}}%
}
\newtcolorbox{promptbox}[1][]{%
  breakable,
  enhanced,
  colback=white,
  colframe=black!18,
  boxrule=0.35pt,
  arc=1pt,
  left=7pt,
  right=7pt,
  top=6pt,
  bottom=6pt,
  fonttitle=\bfseries\footnotesize,
  coltitle=black,
  title={#1},
  before skip=6pt,
  after skip=6pt
}

\usepackage{amsthm}
\usepackage{enumitem}
\usepackage{listings}
\newcommand{\gold}[1]{\colorbox{yellow!25}{\texttt{#1}}}
\newcommand{\pred}[1]{\colorbox{red!15}{\texttt{#1}}}

\newcommand{\benchmark}[0]{\textsc{ScriptIOC-bench}}
\newcommand{\lzero}[0]{Direct}
\newcommand{\lone}[0]{Fixed-transform}
\newcommand{\ltwo}[0]{Program-dependent}

\definecolor{amethyst}{rgb}{0.6, 0.4, 0.8}

\definecolor{darkgreen}{RGB}{0,90,0}

\definecolor{claudeblue}{RGB}{0, 82, 204}   
\newcommand{\claude}[1]{{\color{brown}#1}}

\newcommand{\crep}[2]{{\color{claudeblue}#1}{\color{brown}#2}}

\definecolor{findingaccent}{RGB}{33, 73, 138}

\newtcolorbox{finding}[1][]{
  enhanced,
  colback=findingaccent!4,
  colframe=findingaccent,
  boxrule=0pt,
  leftrule=2.2pt,
  arc=2pt,
  left=10pt, right=8pt, top=5pt, bottom=5pt,
  fonttitle=\bfseries\sffamily,
  coltitle=findingaccent,
  title=#1,
  attach title to upper={\hspace{0.4em}},
  boxed title style={size=fbox, colback=findingaccent!4, colframe=findingaccent!4, frame hidden},
}

\begin{document}

\date{}

\title{\Large \bf \textsc{ScriptIOC-bench}: A Benchmark for Recognizing Actionable \\ Threat Intelligence from Script-Based Malware using LLMs}

\author{
{\rm Hanna Kim$^{1}$ \quad Jian Cui$^{2}$ \quad Minkyoo Song$^{1}$ \quad Hwanjo Heo$^{3}$ \quad Seungwon Shin$^{1}$}\\[3pt]
{\rm Kimin Lee$^{1,*}$ \quad Xiaojing Liao$^{2,*}$}\\[5pt]
$^{1}$KAIST \quad $^{2}$University of Illinois Urbana-Champaign \quad $^{3}$ETRI
}

\newcommand\blfootnote[1]{%
  \begingroup
  \renewcommand\thefootnote{}\footnote{#1}%
  \addtocounter{footnote}{-1}%
  \endgroup
}

\maketitle
\blfootnote{$^{*}$\,Corresponding authors.}

\begin{abstract}
Script-based malware remains a prevalent attack technique. These scripts often contain indicators of compromise (IOCs) that provide actionable threat intelligence. However, statically recovering such indicators is challenging, as relevant values may be dispersed or transformed within code.
Although large language models (LLMs) have shown promise in security analysis, their ability to recover IOCs from malicious scripts remains underexplored.

We present \benchmark{}, a benchmark for measuring static IOC extraction capability on real-world malicious scripts.
The benchmark comprises 634 manually verified JavaScript, PowerShell, and VBScript malware samples covering four IOC types (URLs, domains, IP addresses, and filesystem artifacts).
We further stratify ground-truth IOCs by recovery level, distinguishing directly exposed indicators from those requiring decoding or reconstruction.
Using this benchmark, we evaluate a broad range of proprietary and open-weight LLMs and show that IOC recovery without execution remains challenging across model scales: the strongest model reaches only 65.4 F1.
To characterize how recovery fails, we introduce a false-positive taxonomy and use it to compare the error profiles of the evaluated models.
We further study two mitigations on a small open-weight model, deterministic string utilities and task-specific adaptation, finding that they provide complementary recovery gains, raise precision, and shift errors toward sample-grounded mismatches.
\end{abstract}

\section{Introduction}
\label{sec:introduction}
Script-based malware is an important and widely used attack technique in real-world intrusions. 
By abusing built-in commands and scripting interpreters, adversaries can flexibly execute malicious logic, retrieve follow-on payloads, and interact with the host environment without relying on custom binaries in the initial stage. Consequently, scripts such as PowerShell and JavaScript often function as primary execution or delivery artifacts in cybersecurity attacks~\cite{mitre_t1059,li2019powershell,cova2010jsand}.
At the same time, these scripts often expose information about attacker-controlled infrastructure and accessed resources, making them a valuable source of indicators of compromise (IOCs)~\cite{nist_ioc}.
IOCs refer to artifacts that indicate malicious activity, including URLs and domains, that can support detection, triage, and threat hunting.
Recovering such indicators is therefore an important goal of malware analysis because it enables defenders to move from the analysis of a single sample to the identification of related infections and the disruption of broader malicious activity~\cite{johnson2016nistsp800150,liao2016iace}.

\noindent \textbf{Static IOC recovery.}
IOC extraction imposes a practical scale constraint.
Large analysis platforms report processing roughly 1.2M previously unseen files per day, making scalable automated analysis nontrivial~\cite{virustotal2024scale}.
Given the large volume of script samples encountered in practice, scalable first-line screening is essential. Static analysis has been widely used for this purpose because it enables script artifacts to be examined without execution.
%
Static IOC recovery, however, is more difficult than searching source code for IOC-shaped strings. Malicious scripts routinely split values across literals, encode them through transformations, or reconstruct them through script-specific computation. Recovering such indicators may therefore require reasoning over dispersed program context. This creates a natural opportunity for large language models (LLMs), whose code-understanding capabilities may enable execution-free analysis beyond conventional pattern matching.
Yet existing LLM-based malware studies largely focus on adjacent tasks such as code deobfuscation or behavioral analysis~\cite{fang2024llmcode,jsdeobsbench2025,cybersoceval2025,maleval2025,ctinexus2025}. 
The capabilities and limitations of current LLMs for static IOC recovery therefore remain under-explored.

\noindent \textbf{Our study.}
Against this backdrop, we formulate the following research questions:
\textbf{RQ1.} How reliably can LLMs recover statically obtainable IOCs from malicious scripts?
\textbf{RQ2.} Which forms of recovery remain challenging, from extracting explicit values to sample-specific reconstruction?
\textbf{RQ3.} How do script characteristics such as length affect IOC extraction reliability?
\textbf{RQ4.} What failure modes emerge across model capabilities, and to what extent can they be mitigated?

Addressing these questions requires a benchmark grounded in real-world scripts, so that evaluation reflects the recovery challenges encountered in practice.
Yet no existing dataset provides ground truth restricted to statically obtainable IOCs, as reports from malware-analysis platforms often combine static and dynamic evidence without clearly distinguishing their provenance.
Moreover, indicator-level annotation should characterize how each IOC can be recovered, since different indicators may require different forms of recovery even within a single script.

We therefore present \textbf{\benchmark{}}, the first benchmark for measuring static IOC extraction from real-world script-based malware. It contains 634 manually verified JavaScript, PowerShell, and VBScript samples covering four actionable IOC types: URLs, domains, IP addresses, and filesystem artifacts. Ground-truth indicators are retained only when their values are fully determined by the script file alone. We further annotate each IOC by its recovery level: \textit{\lzero}, when the value appears explicitly in the source; \textit{\lone}, when it can be recovered using common deterministic transformations; and \textit{\ltwo}, when recovery requires following logic specific to the program. 


\noindent \textbf{Findings.}
Using \benchmark{}, we evaluate open-weight LLMs~\cite{Qwen3_8B,Ministral3,Gemma4,Qwen3_C80B,Qwen3_C490B,Qwen3.6_35B,Llama4} from 8B to 480B parameters alongside frontier proprietary models~\cite{Gemini2.5,GPT5.6}, and identify three main findings.
First, \textit{strong LLMs recover directly exposed IOCs well, but performance drops sharply as recovery requires deeper reconstruction.} \textbf{(RQ1-RQ2; \S\ref{sec:eval}-\S\ref{sec:result_recovery_level})} Although the leading LLMs substantially outperform surface-based extractors, even the strongest model reaches only 65.4 F1.
Frontier models recover up to 91.7\% of \lzero{} IOCs but only 31--46\% of \ltwo{} IOCs, showing that recovery through sample-specific program logic remains particularly challenging.
%
Second, \textit{input complexity can limit reliable IOC recovery.} \textbf{(RQ3; \S\ref{sec:result_code_complexity})} Longer scripts generally make LLMs more likely to produce incomplete or unusable outputs and less likely to recover indicators. Such extremely long and redundant samples are common in script-based malware, making inflated context length a practical obstacle to reliable LLM-based IOC recovery.
Third, \textit{smaller models fail differently, and the resulting failure modes are actionable.} \textbf{(RQ4; \S\ref{sec:fp_analysis}-\S\ref{sec:adaptation})} When recovery becomes difficult, smaller models frequently fall back to IOC examples from the prompt, whereas frontier models more often produce sample-grounded reconstruction errors.
We target these weaknesses in a controlled study on a small open-weight model using deterministic string utilities and task-specific adaptation: tools improve grounding and \lzero{} recovery, while adaptation learns recurring reconstruction patterns and improves \lone{} recovery. Together, the two approaches raise precision from 31\% to 48\% while substantially reducing unsupported predictions.

\noindent \textbf{Contributions.}
Our main contributions are as follows:
\begin{itemize}[leftmargin=1.5em]
    \item We introduce \benchmark{}, a benchmark for static IOC extraction from script-based malware, covering JavaScript, PowerShell, and VBScript with four IOC types over 634 manually verified samples.

    \item We systematically evaluate open-weight and frontier LLMs across recovery difficulty and input complexity, revealing a substantial gap between direct extraction and deeper reconstruction, as well as degraded reliability on long inputs.

    \item We identify distinct failure modes across model scales through a fine-grained false-positive taxonomy, and show that tool augmentation and task-specific adaptation shift a small model's errors toward sample-grounded errors.

\end{itemize}

\noindent\textbf{Artifact and Leaderboard.}
We release our datasets and code, and provide a leaderboard; a screenshot is included in Appendix~\ref{apx:leaderboard}.

\section{Background and Related Work}
\label{sec:background}
\subsection{Script-based Malware and IOC Recovery}
\noindent\textbf{Script-based Malware.}
Script-based malware refers to malicious code delivered as interpreted \textit{scripts}, primarily JavaScript, PowerShell, and VBScript. 
These scripts commonly serve as the initial-access or early execution stage in multi-step attacks.
By abusing built-in interpreters and trusted system binaries, adversaries can execute commands, download payloads, and prepare follow-on activity while reducing their reliance on custom binaries~\cite{barrsmith2021survivalism}.
The operational impact is significant.
Campaigns involving script-based loaders such as Emotet, Qakbot, and TrickBot have infected millions of hosts and enabled ransomware payouts in the hundreds of millions of dollars~\cite{doj2021emotet,doj2023qakbot,doj2023trickbot,state2022conti}. 
In 2024, 79\% of detections were malware-free, reflecting the continued shift toward script- and living-off-the-land--based intrusions~\cite{crowdstrike2025gtr}.
To evade detection, script-based malware is often heavily obfuscated~\cite{mitre_t1027} using string encoding, control-flow flattening, and runtime evaluation, such as \texttt{eval} or \texttt{Invoke-Expression}~\cite{xu2012power,bohannon2017revoke,fass2019hidenoseek}.

\noindent\textbf{Indicators of Compromise.}
The primary actionable threat intelligence exposed by these scripts is \textit{indicators of compromise} (IOCs): URLs, domains, IP addresses, and filesystem artifacts that reveal attacker-controlled infrastructure or host-side traces~\cite{johnson2016nistsp800150,paloalto_iocs}. 
Recovering such indicators helps defenders connect a single sample to related infections and disrupt broader campaigns~\cite{liao2016iace,husari2017ttpdrill}. 
However, IOCs in obfuscated scripts rarely appear as plain literals.
Instead, they may be split across strings, encoded, guarded by conditional logic, or reconstructed through runtime-evaluated expressions.
As a result, IOC recovery often requires reasoning over dispersed code context rather than applying surface-level pattern matching alone.

\noindent\textbf{IOC Extraction.}
IOC extraction is commonly approached through dynamic or static analysis.
Dynamic analysis executes a script in an instrumented sandbox to observe runtime artifacts such as resolved URLs, dropped files, and executed commands~\cite{jamalpur2018dynamic,boxjs,li2024powerpeeler,ugarte2019powerdrive}.
Although effective, sandboxing is costly at scale and can be evaded by anti-analysis techniques such as virtualization detection and time-delayed execution~\cite{chen2008antivirt,lindorfer2011envsensitive,afianian2020evasion}.
Static analysis provides an execution-free alternative that scales to large corpora and remains applicable when a script refuses to execute.
Prior work has used abstract syntax trees (ASTs), lexical features, and AST rewriting for script detection and deobfuscation~\cite{bohannon2017revoke, li2019powershell, herrera2020safedeobs, jsimplifier2026}, but reliable IOC extraction from heavily obfuscated scripts remains difficult.
LLMs have recently emerged as an execution-free alternative, given their strong code-reasoning capability.
However, existing LLM-based studies mainly target adjacent tasks such as deobfuscation~\cite{fang2024llmcode, jsdeobsbench2025} and behavior auditing~\cite{cybersoceval2025, maleval2025, cama2025}, or address only a single campaign~\cite{patsakis2024llms}. 
Static IOC extraction across multi-language scripts and campaigns therefore remains largely under-explored.

\subsection{Related Work}
\label{sec:related}
\noindent\textbf{Script-based Malware Analysis.}
Early work on script-based malware analysis applies static and dynamic program analysis either to detect that a script is obfuscated or to recover its underlying semantics. 
For PowerShell, Revoke-Obfuscation~\cite{bohannon2017revoke} treats obfuscation as binary classification, training logistic regression over AST-derived statistical features.
In an effort to recover the underlying semantics, the literature successively focuses on finer program units:
Li~et~al.~\cite{li2019powershell} emulate individual AST subtrees in isolation; PowerDrive~\cite{ugarte2019powerdrive} hooks the PowerShell interpreter's evaluation cmdlets at runtime;
and PowerPeeler~\cite{li2024powerpeeler} traces execution at the instruction level.
Similarly, for JavaScript, SAFE-Deobs~\cite{herrera2020safedeobs} applies static compiler optimizations such as constant folding and dead-code elimination, and JSimplifier~\cite{jsimplifier2026} adds dynamic tracing and LLM-based identifier renaming across 44K real samples.

A more recent line applies LLMs directly to malicious-script reasoning. Fang~et~al.~\cite{fang2024llmcode} systematically evaluate frontier LLMs on general code understanding and JavaScript deobfuscation. 
JsDeObsBench~\cite{jsdeobsbench2025} establishes a focused benchmark for evaluating LLMs' capability in deobfuscating JavaScript, under zero- and few-shot prompting. 
CyberSOCEval~\cite{cybersoceval2025}, MalEval~\cite{maleval2025}, and Cama~\cite{cama2025} benchmark prompted LLMs on SOC-style threat reasoning, fine-grained malware behavior auditing, and Android function identification, respectively.
However, all of these efforts target outputs other than direct actionable IOCs (deobfuscated code, behavior summaries, etc.), and they also evaluate only vanilla LLMs on a single language family.

\noindent\textbf{Threat Intelligence Extraction.}
Early studies on threat intelligence extraction leverage classical NLP approaches: iACE~\cite{liao2016iace} and TTPDrill~\cite{husari2017ttpdrill} extract OpenIOC items and structured TTPs from technical articles via stable grammatical relations and information-retrieval techniques.
Recent studies focus on LLM-driven pipelines~\cite{ctinexus2025, action2023}: they construct CTI knowledge graphs from natural-language reports and ingest real-world CTI feeds at scale. 
Closest to our setting, Patsakis~et~al.~\cite{patsakis2024llms} evaluate frontier LLMs on PowerShell IOC extraction over a single Emotet campaign, with the best-performing model reaching 69.6\%/88.8\% accuracy on URLs/domains.
Our work, however, provides a comprehensive benchmark built from real-world malicious scripts across three script languages and four IOC types.
We systematically evaluate model performance across recovery difficulty and conduct a fine-grained analysis of their failure modes.

\subsection{Problem Scope}
\label{sec:scope}

This work studies \emph{static IOC extraction from script-based malware using LLMs}. The analysis is restricted to static reasoning over the script source.
We define an IOC as \emph{statically recoverable} when its value is fully determined by the script and embedded data, without execution or runtime input. This includes values obtained through decoding, string reconstruction, constant propagation, or bounded deterministic computation. 
Some recoverable IOCs may still require interpreting program-specific logic beyond fixed transformations.
We focus on three of the widely used scripting languages: JavaScript, PowerShell, and VBScript; and four actionable IOC categories: URLs, domains, IP addresses, and filesystem artifacts.
\section{\benchmark}
\begin{figure}[t]
    \centering
    \includegraphics[width=1\linewidth]{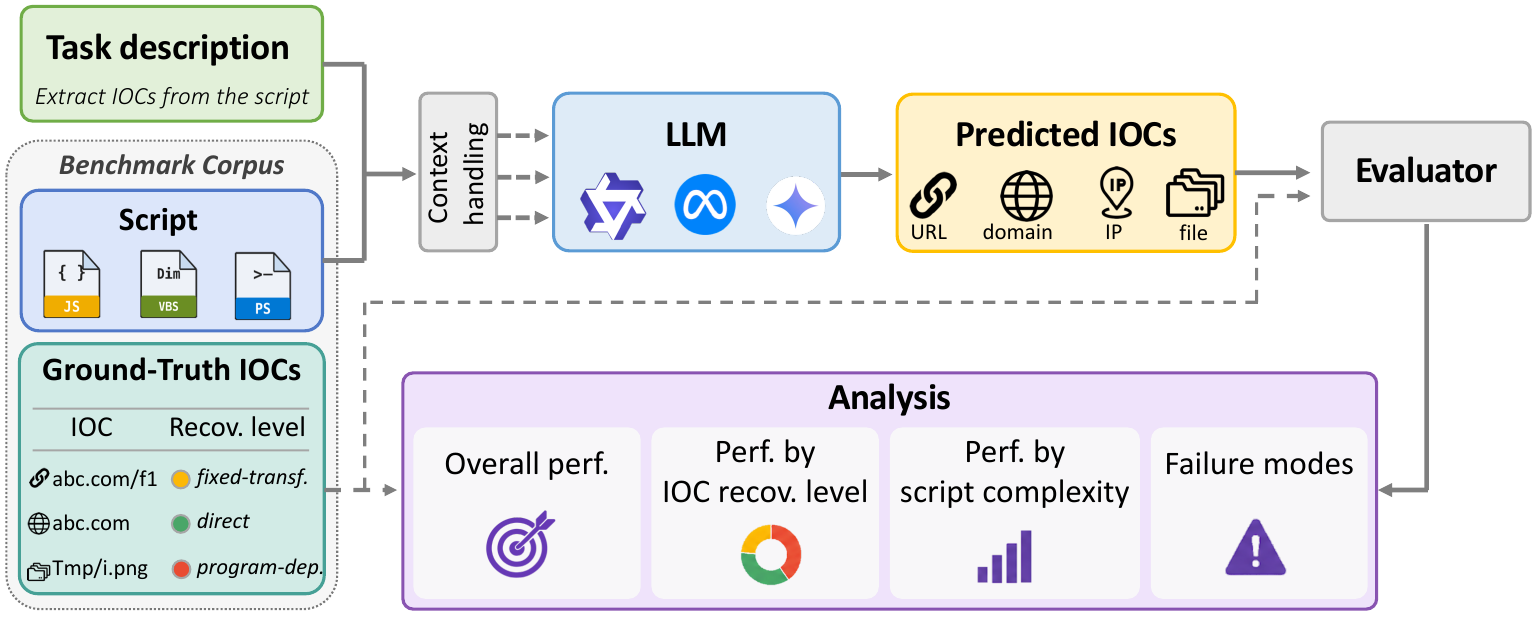}
    \caption{Overview of \benchmark{}. Given a task description and a malicious script, an LLM extracts statically obtainable IOCs, which are evaluated against ground-truth IOCs.
    We analyze model performance along four dimensions corresponding to our research questions: overall performance (RQ1), IOC recovery level (RQ2), input complexity (RQ3), and failure modes (RQ4).
    }
    \label{fig:overview}
\end{figure}

In this section, we present \benchmark, a benchmark for \emph{static} IOC extraction from script-based malware.
Figure~\ref{fig:overview} presents an overview of our framework.

\noindent \textbf{Overview.}
The framework consists of three main components: task input, LLM-based IOC extraction, and evaluation.
\benchmark{} provides a corpus of \emph{labeled} malicious script samples spanning JavaScript, PowerShell, and VBScript, where each instance pairs a script with ground-truth IOC labels, each further annotated with a \textit{recovery level} (\S\ref{sec:benchmark-evaluation}).
Given a script and task description, the LLM produces IOC predictions, which are scored against the ground truth, with further analysis by recovery level, input complexity, and false-positive profiles.

The remainder of this section details these components: the task and its input corpus (\S\ref{sec:task-formulation}, \S\ref{sec:benchmark-composition}), the inference setup (\S\ref{sec:inference-setup}), and the evaluation framework (\S\ref{sec:benchmark-evaluation}).

\subsection{Task Formulation}
\label{sec:task-formulation}
\noindent \textbf{Task Definition.}
For each input, the model is given a natural-language task description and a malicious script sample.
The model is instructed to statically extract IOCs from the script without executing it and is restricted to output a structured set of indicators. 
Since such indicators are often obfuscated, encoded, or dispersed across the code, successful extraction requires reconstructing them from script-level context.
The task can be formally defined as follows:
Given an input \(x=(d,s)\), where \(d\) denotes the task description and \(s\) denotes the script sample, the model outputs a set of typed IOCs:
\[
\hat{Y} = f_\theta(d,s).
\]
Each predicted IOC is represented as a pair \((v,t)\), where \(v\) is the extracted IOC value and its IOC type, \(t \in \{\textsc{url}, \textsc{domain}, \textsc{ip}, \textsc{file}\}\).




\subsection{Benchmark Composition}
\label{sec:benchmark-composition}

\noindent \textbf{Target Script Languages.}
We focus on JavaScript, PowerShell, and VBScript, three widely used scripting languages in real-world malware~\cite{redcanary_tdr, hpwolf2024sept}. 
These three languages capture distinct but complementary features under different script-execution settings relevant to real-world malicious activity.

\noindent $\bullet$ \textit{JavaScript (JS)} is a general-purpose interpreted scripting language that operates across browsers and other runtime environments. 
In malware, its portability and flexibility make it suitable for lures, lightweight loaders, and staged payload delivery across multiple execution contexts.

\noindent $\bullet$ \textit{PowerShell (PS)} is a command-line shell and scripting language designed for system administration and automation, particularly in Windows environments. 
Because it is natively available on many Windows systems and provides access to system functionality, it is frequently abused for command execution, payload staging, and post-compromise activity while blending with legitimate administrative behavior.

\noindent $\bullet$ \textit{VBScript (VBS)} is a lightweight Visual Basic--derived scripting language commonly executed through Windows Script Host. 
Despite being an old Windows component, it remains relevant in malware because it is still available on modern Windows systems and can be used as a simple loader or dropper to launch additional tools.
Recent malware campaigns have also continued to use VBScript as an initial delivery or execution component~\cite{vbs_whatsapp}.


\noindent \textbf{Target IOC Types.}
In this work, we define an IOC as any maliciously relevant artifact recoverable from the script through \textit{static analysis}, including artifacts embedded as literals and those obtained through static decoding, deobfuscation, or string reconstruction.

We focus on four IOC types: URLs, domains, IP addresses, and filesystem artifacts. 
These categories serve as actionable threat intelligence, capturing different aspects of malicious activity, such as network infrastructure and host-side traces.
Furthermore, they usually appear in script-based malware: in payload download commands, command-and-control references, persistence logic, and file-write operations.

For example, a script-based downloader may embed a URL or domain used to retrieve a secondary payload, or a hard-coded IP address used to contact attacker-controlled infrastructure. 
It may also specify a file path in a temporary directory or user profile to store the downloaded payload or maintain persistence. 
Even when such artifacts are split across multiple strings, encoded, or lightly obfuscated, they are often recoverable through purely static analysis, making them suitable targets for extraction.

\noindent \textbf{Dataset.}
The benchmark data contains 634 manually verified malicious script samples across the three target languages.
Samples are drawn from two complementary sources: recent scripts collected from Filescan.io~\cite{filescanio} (Nov 2025--Mar 2026) covering JS, PS, and VBS; and historical JavaScript samples from the Hynek Petrak collection (2015--2017)~\cite{HynekPetrak}, which form a separate subset capturing earlier obfuscation styles.
Two security experts construct the ground truth by direct static analysis of each script, keeping only statically recoverable indicators.
Each ground-truth IOC is further annotated with an \emph{IOC recovery level}---\textit{\lzero{}}, \textit{\lone{}}, or \textit{\ltwo{}}---according to whether its value appears verbatim, requires a common deterministic transformation, or must be reconstructed from program-specific logic.
Table~\ref{tab:benchmark_dataset} summarizes the composition by source, time period, and language.
Detailed sample collection, annotation protocol, and corpus-level analysis are presented in \S\ref{sec:dataset}.

\subsection{LLM Inference Setup}
\label{sec:inference-setup}
LLM serves as the inference component of \benchmark{}. Given a task description and a malicious script, the model identifies IOCs from the script, including values that may require resolving indirect references or deobfuscation. The model response is parsed into a structured set of predicted IOCs for evaluation.

\noindent\textbf{Context handling.} Many malicious script samples exceed the context window of typical LLMs (\S\ref{sec:dataset-analysis}), and a single-pass prompt would truncate the input and discard security-relevant code.
To address this, we use a standard map--reduce-style procedure for such cases.
Scripts that fit within the context window are processed directly; longer scripts are split into token-bounded chunks, processed independently, and their intermediate findings are subsequently aggregated to produce the final output.

\subsection{Evaluation Framework}
\label{sec:benchmark-evaluation}

\noindent \textbf{Metrics.}
The LLM's capability of extracting IOCs is evaluated on per-IOC precision, recall, and F1 score, computed independently for each IOC type; the detailed matching rules are described in \S\ref{sec:metrics}.

We evaluate models along four diagnostic axes.

\noindent
\hangindent=1.2em
\hangafter=1
\textbullet\hspace{0.5em}\noindent\textbf{Overall performance (RQ1).}
We report aggregate precision, recall, and F1, and compare against regex-based baselines to assess the overall reliability of LLM-based IOC recovery.
  
\noindent
\hangindent=1.2em
\hangafter=1
\textbullet\hspace{0.5em}\noindent\textbf{IOC recovery level (RQ2).}
Different recovery levels place different demands on the model, so we analyze performance across \lzero{}, \lone{}, and \ltwo{} indicators to identify which forms of recovery remain challenging.

\noindent
\hangindent=1.2em
\hangafter=1
\textbullet\hspace{0.5em}\noindent\textbf{Input complexity (RQ3).}
To assess robustness under increasingly demanding inputs, we analyze extraction quality across the input-size and code-redundancy measures.

\noindent
\hangindent=1.2em
\hangafter=1
\textbullet\hspace{0.5em}\noindent\textbf{Failure modes (RQ4).}
Different failure modes impose different verification costs and call for different mitigations, so we characterize false positives using the taxonomy and examine how error types vary across models.

\section{Datasets}
\label{sec:dataset}

This section details the construction and analysis of the \benchmark{} dataset.
\S\ref{sec:dataset-construction} describes how malware samples were collected and how their IOC labels were verified under category-specific annotation rules.
In \S\ref{sec:dataset-analysis}, we detail the analysis of the resulting dataset along two aspects: input complexity and IOC label distribution.

\subsection{Dataset Construction}
\label{sec:dataset-construction}

\noindent\textbf{Malicious Script Collection.}
We collect malware samples from two sources, Filescan.io~\cite{filescanio} and Hynek Petrak~\cite{HynekPetrak}, spanning Nov. 2025--Mar. 2026 and 2015--2017, respectively.
Specifically, we first collected newly-observed JS, PS, and VBS malware samples from Filescan.io~\cite{filescanio}, a malware database that stores analyzed samples and assigns quality labels, e.g., \textit{confirmed threat}. We randomly sampled scripts marked as \textit{high risk} or \textit{confirmed threat} that exposed at least one target IOC and were available for download.

We additionally incorporated malicious JavaScript samples from the JS Malware Collection released by Hynek Petrak~\cite{HynekPetrak}. This collection contains real-world JavaScript malware samples from 2015 to 2017. We randomly sampled a subset of this collection to expand the JavaScript portion of the dataset. Details of the dataset are summarized in Table~\ref{tab:benchmark_dataset}.

\begin{table}[t]
\centering
\caption{Dataset composition by source, collection period, and script language.}
\footnotesize
{\setlength{\tabcolsep}{3.5pt}
\begin{tabular}{l c c r l}
\toprule
Source & Period & Lang. & \# Samples & Dataset\\
\midrule
\multirow{3}{*}{Filescan.io}
& Nov. 2025--Feb. 2026 & JS  & 97 &  JS$_f$\\
& Jan. 2026--Mar. 2026 & PS & 75  & PS$_f$\\
& Jan. 2026--Mar. 2026 & VBS  & 79  & VBS$_f$\\
\midrule
Hynek Petrak
& 2015--2017 & JS & 383 & JS$_{hp}$\\
\bottomrule
\end{tabular}
}

\label{tab:benchmark_dataset}
\end{table}


\noindent\textbf{IOC Annotation.}
Evaluating static IOC extraction requires that every labeled indicator actually be recoverable from the script alone.
Since no existing source provides labels curated for static recoverability, by following established annotation process~\cite{votipka2020understanding, mcdonald2019reliability},  we construct the ground truth through expert manual static analysis. 
Two security experts manually analyze each script, deobfuscating when necessary, to identify and verify each IOC directly from the source.
The ground truth retains only \emph{statically obtainable} indicators: values fully determined by the script and its embedded data.
Static recoverability does not require cleartext presence: retained indicators may require multi-step deobfuscation, but values that depend on runtime or external state are excluded.

The annotation proceeds in three stages.
The experts first independently review a random subset of 32 scripts (about 5\% of the 634 samples) to identify statically recoverable IOCs and develop initial annotation guidelines. The guidelines specify the annotation scope and category-specific decision rules, including inclusion and exclusion criteria for each IOC type. Using these initial guidelines, they independently annotate another random subset of 32 scripts, compare their annotations, resolve disagreements, and refine the guidelines to clarify ambiguous cases. They then annotate the remaining samples under the finalized guidelines to obtain the final ground-truth IOC set.

Throughout annotation, IOC reports from four malware analysis platforms (Filescan.io, VirusTotal~\cite{virustotal}, Triage~\cite{triage}, and Hybrid Analysis~\cite{hybridanalysis}) were consulted only as auxiliary references for cross-checking candidate indicators: such reports mix statically observable artifacts with dynamic-only results and platform-specific enrichment, and may miss indicators when malware evades sandbox execution.
Retained artifacts follow the category-specific annotation rules below.

For URLs, we annotate both fully specified URLs and URLs without a scheme when a host and path are explicitly present in the script.
URLs hosted on benign shared services (e.g., \texttt{api.telegram.org}) are retained in the URL-level ground truth when they are presumed to be used maliciously, such as for payload delivery.
Conversely, URLs that are themselves benign, e.g., pointing to legitimate service content rather than attacker-staged resources, are not added; benignness is assessed using VirusTotal~\cite{virustotal}.
For domains, those belonging to these abused legitimate services are excluded from the domain-level ground truth because they are not attacker-controlled, even when URLs hosted on them are retained at the URL level.
For filesystem artifacts, we define an artifact as a file or directory on the host that is the statically identifiable target of a creation, drop, extraction, copy, move, or rename operation in the script. These include full and partial paths, as well as file or directory names whose parent paths cannot be determined. Indicators reconstructed through constant propagation or string concatenation are also included. Environment-variable references such as \texttt{\%TEMP\%} and \texttt{\%APPDATA\%} are retained in symbolic form rather than resolved to machine-specific paths.

\noindent\textbf{IOC Recovery Levels.}
\label{sec:recovery-levels}
As a final labeling step, we assign each ground-truth IOC a recovery level to identify where models succeed and where they struggle in IOC extraction.
A model may recover an IOC simply by reading a surface string, by undoing a common encoding, or by reasoning through script-specific logic; aggregate extraction results do not distinguish among these qualitatively different cases. 
We therefore assign each IOC one of three levels based on how it can be recovered from the script: \textit{\lzero{}}, indicators that appear verbatim in the source and are recoverable by surface-form matching; \textit{\lone{}}, indicators hidden behind common, parameter-free transforms that a fixed decoder can invert; and \textit{\ltwo{}}, indicators whose recovery requires following logic specific to the script, such as how a string is assembled or decoded.

The assignment uses a fixed decoder that applies a set of common reversible transforms over surface strings or bounded parameters, never executing the script. The transforms fall into two families: \emph{reassembly} operators, which only reorder characters already present in the source (literal concatenation, split/join, etc.), and \emph{transcode} operators, which invert a codec (Base64 decoding, character-code arrays, etc.).
The details and their selection rationale are described in Appendix~\ref{apx:recovery-levels}.

Recovery levels are assigned to individual IOCs rather than entire scripts, allowing a single script to contain indicators at multiple levels. For example, a script may contain a \lzero{} domain appearing verbatim alongside a \ltwo{} URL reconstructed through script-specific logic. 
Note that \ltwo{} denotes indicators that are not recovered by the fixed transform set, rather than indicators that are inherently unrecoverable through static analysis.

\subsection{Dataset Analysis}
\label{sec:dataset-analysis}
In this section, we characterize the benchmark at two levels: sample-level input complexity (\S\ref{sec:dataset_complexity}) and IOC-level distribution and recovery composition (\S\ref{sec:ioc_level_analysis}).

\subsubsection{Input Complexity}
\label{sec:dataset_complexity}
We measure the complexity of the benchmark along two complementary dimensions: code token length and gzip compression ratio.
\textbf{Code token length} measures the size of each script in model tokens. In the LLM setting, longer scripts are generally more difficult to analyze because they occupy a larger fraction of the available context window, making it harder to aggregate relevant evidence across the full input.
\textbf{Gzip compression ratio} measures how compressible a script is. Lower ratios
indicate more compressible code, which is generally associated with
greater repetition, whereas higher ratios indicate lower
compressibility. We use this measure to capture structural variation
that is not reflected by input length alone.


\begin{figure}[t]
    \centering

    \begin{subfigure}{1\linewidth}
        \centering
        \includegraphics[width=\textwidth]{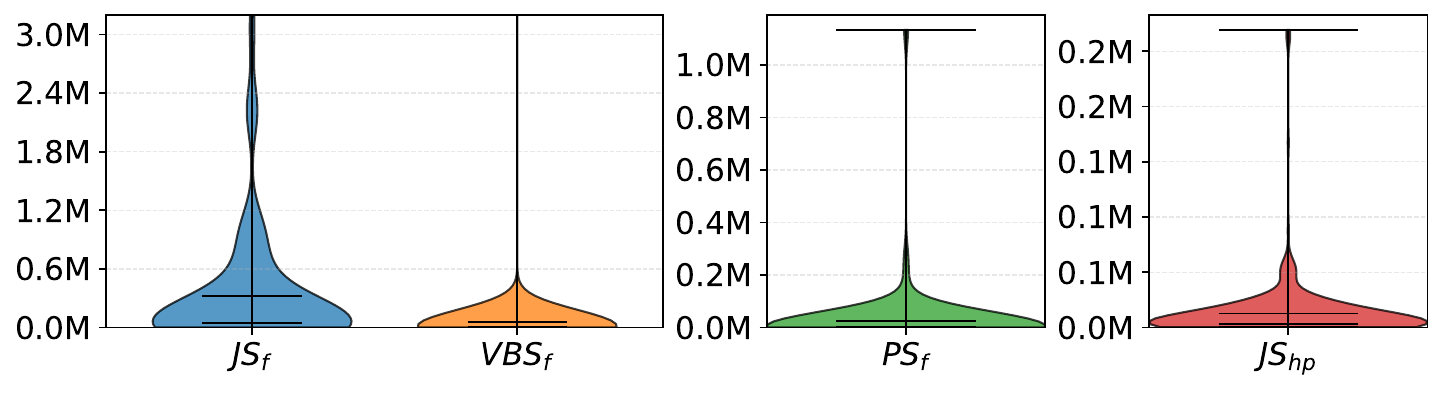}
        \vspace{-7mm}
        \caption{Code token length}
        \label{fig:tok_len}
    \end{subfigure}

    \begin{subfigure}{\linewidth}
        \centering
        \includegraphics[width=\textwidth]{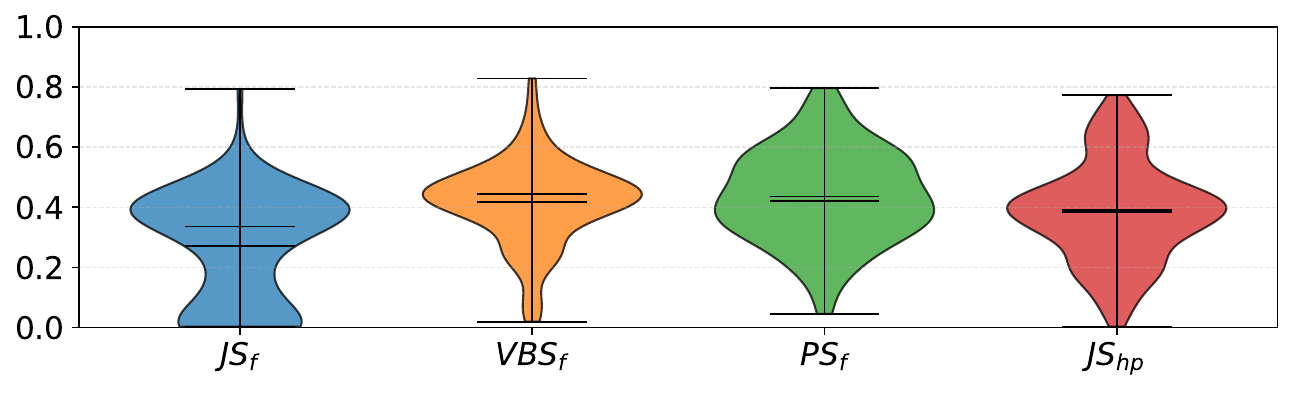}
        \vspace{-7mm}
        \caption{Gzip compression ratio}
        \label{fig:str_entropy}
    \end{subfigure}

    \caption{Distributions of (a) code token length and (b) gzip compression ratio across the benchmark dataset.}
    \label{fig:code_complexity}
\end{figure}

Figure~\ref{fig:code_complexity} shows the distribution of complexity metrics across benchmark subsets. Malware samples exhibit heavy-tailed token-length distributions, with $JS_f$ markedly longer than the other subsets. Mean token lengths are 325.6K, 25.4K, and 58.0K for $JS_f$, $PS_f$, and $VBS_f$, respectively, while $JS_{hp}$ averages 13.3K tokens. Although shorter on average, $PS_f$ and $VBS_f$ also retain long upper tails. Such input lengths can exceed the practical capacity of many widely used models and remain demanding even for frontier long-context models\footnote{Google’s Gemini family has supported 1M-token context windows as standard for long-context use.}. Therefore, scripts of this size can still place meaningful pressure on long-context reasoning and evidence aggregation.

The gzip compression ratios span a wide range across all datasets, indicating diversity in structural redundancy among samples. 
This diversity may pose challenges for LLM-based extraction in two different ways. Low-ratio samples can dilute genuine IOCs within large volumes of repetitive filler code, potentially degrading the signal-to-noise ratio within the model's context. High-ratio samples, on the other hand, may conceal IOCs inside high-entropy encoded content (Base64 blobs, XOR strings), which could require the model to recognize and reverse encoding schemes — a capability that LLMs are known to perform unreliably under stacked obfuscation~\cite{jsdeobsbench2025}.

Together, these results indicate that benchmark difficulty arises from different sources across subsets, including input length and structural irregularity.

\ifdefined\lvlunit\else\newlength{\lvlunit}\fi
\setlength{\lvlunit}{0.42pt}
\definecolor{lvlzero}{RGB}{201,206,212}
\definecolor{lvlone}{RGB}{93,139,174}
\definecolor{lvltwo}{RGB}{178,60,48}
\providecommand{\lvlbar}[3]{%
  {\color{lvlzero}\rule[-1.2pt]{#1\lvlunit}{5.5pt}}%
  {\color{lvlone}\rule[-1.2pt]{#2\lvlunit}{5.5pt}}%
  {\color{lvltwo}\rule[-1.2pt]{#3\lvlunit}{5.5pt}}}
\providecommand{\lvlrow}[3]{\lvlbar{#1}{#2}{#3}\;{\scriptsize #1/#2/#3}}
\providecommand{\lvlbox}[1]{{\color{#1}\rule[-0.5pt]{5pt}{5pt}}}

\begin{table}[t]
\centering
\caption{IOC distribution and recovery-level composition across script categories. Columns report prevalence, total and unique IOC counts, average instances per sample, and the share of instances at each recovery level: L0 = \lvlbox{lvlzero}\,\lzero{}, L1 = \lvlbox{lvlone}\,\lone{}, and L2 = \lvlbox{lvltwo}\,\ltwo{}.}
\footnotesize
\setlength{\tabcolsep}{2.5pt}
\renewcommand{\arraystretch}{1.1}
\begin{tabular}{l l r r r l}
\toprule
Dataset & Type & \% Pos. & Inst. (Uniq.) & Avg. & Rec. dist. (L0/L1/L2) \\
\midrule
\multirow{4}{*}{JS$_f$}
  & URL      & 99.0 & 110 (104) & 1.13 & \lvlrow{56}{3}{41} \\
  & Domain   & 57.7 & 63 (46) & 0.65 & \lvlrow{44}{5}{51} \\
  & IP       & 18.6 & 18 (11) & 0.19 & \lvlrow{78}{0}{22} \\
  & File     & 6.2 & 10 (10) & 0.10 & \lvlrow{40}{30}{30} \\
\midrule
\multirow{4}{*}{PS$_f$}
  & URL      & 92.0 & 147 (140) & 1.96 & \lvlrow{65}{13}{22} \\
  & Domain   & 64.0 & 66 (58) & 0.88 & \lvlrow{53}{17}{30} \\
  & IP       & 21.3 & 16 (14) & 0.21 & \lvlrow{56}{6}{38} \\
  & File     & 30.7 & 46 (44) & 0.61 & \lvlrow{81}{14}{5} \\
\midrule
\multirow{4}{*}{VBS$_f$}
  & URL      & 98.7 & 140 (100) & 1.77 & \lvlrow{34}{16}{50} \\
  & Domain   & 81.0 & 86 (52) & 1.09 & \lvlrow{34}{11}{55} \\
  & IP       & 13.9 & 12 (10) & 0.15 & \lvlrow{25}{67}{8} \\
  & File     & 26.6 & 35 (33) & 0.44 & \lvlrow{70}{15}{15} \\
\midrule
\multirow{4}{*}{JS$_{hp}$}
  & URL      & 100.0 & 941 (785) & 2.46 & \lvlrow{25}{19}{56} \\
  & Domain   & 99.7 & 781 (561) & 2.04 & \lvlrow{30}{22}{48} \\
  & IP       & 0.5 & 2 (2) & 0.01 & \lvlrow{0}{50}{50} \\
  & File     & 81.5 & 417 (264) & 1.09 & \lvlrow{9}{15}{76} \\
\bottomrule
\end{tabular}
\label{tab:ioc_label_distribution}
\end{table}


\subsubsection{IOC Label Distribution}
\label{sec:ioc_level_analysis}
Here, we characterize IOC-type prevalence and recovery-level composition across script subsets (Table~\ref{tab:ioc_label_distribution}) and examine the differences reflected in these distributions.

\noindent\textbf{IOC distribution and behavioral patterns.}
Across all script categories, URLs and domains are the predominant IOC types, while the relative distribution of IOC labels varies substantially by script type, reflecting distinct malware behaviors.
The recent JS subset, $JS_f$, has an IOC profile that is sparse and URL-centric, typically involving a single primary endpoint, with direct IP communication appearing only occasionally.

The PS subset, $PS_f$, has the broadest URL distribution and is more consistent with multi-stage droppers.
These samples often contact multiple URLs, sometimes using several paths on the same host, as they retrieve components such as launchers, DLLs, archives, or follow-on scripts from shared infrastructure. Another notable characteristic is the frequent use of legitimate platforms and APIs, including GitHub, Dropbox, and Telegram, as storage or delivery channels.


The VBS subset, $VBS_f$, follows a distinct two-stage structure. Many VBS samples exhibit a regular two-URL pattern, with 42\% of samples containing an initial retrieval URL and a second-stage payload URL. Pastebin-like services are also frequently used to host obfuscated script content for later execution. Compared with PS, VBS samples less often rely on many endpoints from the same server and more often follow this simpler staged-delivery pattern. 

Finally, the historical JS subset ($JS_{hp}$) is particularly dense in domains. This largely reflects downloader-style scripts that iterate over multiple candidate domains and repeatedly issue HTTP requests until a valid payload source is found. As a result, even relatively compact scripts can generate rich domain annotations by embedding multiple fallback hosts within the same retrieval logic.


\noindent\textbf{Recovery-level composition.}
The recovery-level composition reveals differences that are not apparent from IOC frequency alone. $PS_f$ is comparatively dominated by directly observable indicators, particularly for file IOCs, whereas $JS_{hp}$ contains the largest share of \ltwo{} recovery. This contrast also varies by IOC type: files are largely \lzero{} in $PS_f$ and $VBS_f$, while URLs and domains more often require reconstruction from program logic. $JS_{hp}$ exhibits the opposite extreme for files, with 76\% requiring \ltwo{} recovery. \lone{} indicators form a relatively small share overall, suggesting that many non-direct IOCs cannot be recovered by applying a fixed set of transformations alone.

\begin{table}[t]
\centering
\caption{LLM configurations used for evaluation.}
\footnotesize
\begin{threeparttable}
{\setlength{\tabcolsep}{2pt}
\resizebox{\columnwidth}{!}{%
\begin{tabular}{l r l r}
\toprule
Model & Size & Model ID & Context \\
\midrule
Qwen3-8B~\cite{Qwen3_8B}              & 8.2B  & Qwen3-8B                           & 32K  \\
Ministral3-8B~\cite{Ministral3}         & 8B    & Ministral-3-8B-Instruct-2512       & 256K \\
Gemma4-8B~\cite{Gemma4}             & 8B    & Gemma-4-E4B-it                     & 128K \\
Qwen3-C80B$^{*}$~\cite{Qwen3_C80B}      & 80B   & Qwen3-Coder-Next                   & 262K \\
Qwen3-C480B$^{*}$~\cite{Qwen3_C490B}     & 480B  & Qwen3-Coder-480B-A35B-Instruct     & 262K \\
Qwen3.6-35B~\cite{Qwen3.6_35B}  & 35B   & Qwen3.6-35B-A3B          & 262K \\
Llama4-S109B~\cite{Llama4}              & 109B  & Llama-4-Scout-17B-16E-Instruct     & 10M  \\
\midrule
Gemini-Pro~\cite{Gemini2.5}            & /     & gemini-2.5-pro                     & 1M   \\
Gemini-Flash~\cite{Gemini2.5}          & /     & gemini-2.5-flash                   & 1M   \\
GPT-Terra~\cite{GPT5.6} & /  & gpt-5.6-terra            & 922K \\
\bottomrule
\end{tabular}%
}
}

\begin{tablenotes}[flushleft]
\footnotesize
\item \textit{Note.} $^{*}$ indicates models specialized for code generation.
\end{tablenotes}
\end{threeparttable}

\label{tab:llm_selection}
\end{table}

\section{Experiments}
\subsection{LLM Selection}
As shown in Table~\ref{tab:llm_selection}, we evaluate ten models spanning a wide range of scales, from 8B open-weight models to 480B and frontier proprietary models, to examine how model scale and family affect IOC extraction performance.
We focused mainly on open-weight models because they provide greater transparency, reproducibility, and flexibility for controlled evaluation. 
Smaller models also allow us to assess whether lightweight LLMs can remain competitive under practical constraints such as limited compute, memory, and latency budgets. 
We included code-specialized LLMs alongside general-purpose models, as they are expected to exhibit strong code understanding capabilities. 
We also incorporated three frontier proprietary models---Gemini-Pro, Gemini-Flash, and GPT-Terra (collectively referred to as \emph{frontier models})---as reference baselines.
Note that Claude Opus 4.6 and 4.7 were also tested but excluded because they returned \texttt{stop\_reason: "refusal"} on the evaluation samples.
Details of the selected LLM configurations are provided in Table~\ref{tab:llm_selection}.

\subsection{Evaluation Setup}
\label{sec:eval-setup}

\subsubsection{Baselines}

We compare LLMs with three rule-based extraction tools: \texttt{iocextract}~\cite{iocextract}, \texttt{URLExtract}~\cite{urlextract}, and \texttt{MSTICPy}~\cite{msticpy}. 
\texttt{iocextract} extracts common indicators such as URLs and IP addresses and additionally supports several obfuscated URL forms, including defanged, URL-encoded, hex-encoded, and Base64-encoded variants. \texttt{URLExtract} is a lightweight URL extractor based on top-level-domain (TLD) matching. 
\texttt{MSTICPy} is a Python library for InfoSec investigation and threat hunting, and its IoCExtract component provides broader regex-based IOC extraction, including URLs, DNS domains, IP addresses, hashes, and file paths.
Together, they provide useful reference points for analyzing the strengths and limitations of LLMs in IOC extraction.
We apply the same post-processing filters to each tool's raw output; details are provided in Appendix~\ref{apx:regex_baselines}.

\subsubsection{Inference Setup}

The extraction prompt instructs the model to emit a three-block structured output: \texttt{<analysis>} containing behavioral reasoning and deobfuscation steps, \texttt{<deobfuscated\_code>} containing an analyst-readable code reconstruction, and \texttt{<answer>} containing the final JSON-encoded IOC set with classification verdict. The prompt template is in Appendix~\ref{apx:prompt}.
 
\noindent\textbf{Long-context handling.}
For samples exceeding the per-chunk code budget derived from each model's served context length, we apply a map--reduce procedure: each chunk is processed independently to record partial findings (deobfuscation notes, IOC candidates, unresolved references), and a final reduce call aggregates these into the structured output. 
Short samples bypass chunking entirely. 
The model serving parameters, chunking budget formula, and per-model triggering statistics are detailed in Appendix~\ref{apx:inference_config} and Appendix~\ref{apx:chunking}.

\begin{table*}[t]
\caption{Main results across models. For each script category and overall, we report the response-failure rate (RFail) and micro-averaged precision (P), recall (R), and F1 over IOC instances. Response failures are counted as empty predictions.}
\centering
\footnotesize
\setlength{\tabcolsep}{3pt}
\resizebox{\textwidth}{!}{%
\begin{tabular}{l  cccc cccc cccc cccc cccc}
\toprule
\multirow{2}{*}{Model}
& \multicolumn{4}{c}{$JS_f$}
& \multicolumn{4}{c}{$PS_f$}
& \multicolumn{4}{c}{$VBS_f$}
& \multicolumn{4}{c}{$JS_{hp}$}
& \multicolumn{4}{c}{Overall} \\
\cmidrule(lr){2-5}
\cmidrule(lr){6-9}
\cmidrule(lr){10-13}
\cmidrule(lr){14-17}
\cmidrule(lr){18-21}
& RFail & P & R & F1
& RFail & P & R & F1
& RFail & P & R & F1
& RFail & P & R & F1
& RFail & P & R & F1 \\
\midrule
Qwen3-8B
& 8.4 & 38.5 & 45.3 & 41.6
& 17.9 & 44.4 & 43.7 & 44.0
& 12.9 & 30.4 & 27.1 & 28.7
& 14.4 & 28.0 & 22.0 & 24.7
& 13.7 & 31.3 & 26.5 & 28.7 \\
Ministral3-8B
& 3.8 & 23.7 & 36.7 & 28.7
& 4.9 & 36.2 & 44.1 & 39.8
& 30.0 & 28.4 & 27.5 & 27.9
& 5.1 & 22.6 & 30.0 & 25.8
& 8.0 & 24.3 & 31.6 & 27.5 \\
Gemma4-8B
& 10.0 & 72.4 & 40.0 & 51.5
& 5.8 & 59.6 & 12.2 & 20.2
& 0.4 & 48.9 & 7.3 & 12.8
& 6.2 & 45.4 & 12.3 & 19.3
& 6.0 & 50.8 & 13.9 & 21.8 \\
Qwen3-C80B$^{*}$
& 11.3 & 58.4 & 46.8 & 51.9
& 8.0 & 45.2 & 54.1 & 49.3
& 34.2 & 44.4 & 34.3 & 38.7
& 36.8 & 54.9 & 35.6 & 43.2
& 29.2 & 52.5 & 38.1 & 44.2 \\
Qwen3.6-35B
& 6.8 & 56.3 & 49.8 & 52.8
& 16.9 & 63.9 & 48.7 & \textbf{55.2}
& 17.3 & 51.5 & 24.9 & 33.6
& 14.4 & 55.9 & 41.3 & 47.6
& 13.9 & 56.5 & 41.0 & 47.5 \\
Llama4-S109B
& 4.6 & 45.5 & 32.7 & 38.0
& 2.2 & 42.9 & 46.5 & 44.6
& 4.3 & 39.5 & 24.9 & 30.5
& 3.0 & 39.1 & 27.6 & 32.3
& 3.3 & 40.1 & 29.5 & 34.0 \\
Qwen3-C480B$^{*}$
& 3.8 & 55.3 & 53.9 & \textbf{54.5}
& 0.0 & 52.1 & 57.1 & 54.5
& 0.0 & 41.3 & 41.3 & \textbf{41.3}
& 0.0 & 53.2 & 47.0 & \textbf{49.9}
& 0.6 & 51.9 & 47.9 & \textbf{49.8} \\
\midrule
Gemini-Pro
& 17.5 & 60.2 & 55.7 & 57.9
& 7.8 & 51.9 & 58.1 & 54.8
& 4.9 & 48.0 & 59.9 & 53.3
& 1.8 & 68.9 & 69.6 & \textbf{69.2}
& 5.3 & 63.7 & 66.4 & 65.0 \\
Gemini-Flash
& 56.1 & 74.7 & 33.5 & 46.3
& 16.9 & 46.3 & 54.7 & 50.2
& 48.1 & 64.1 & 45.1 & 53.0
& 14.1 & 72.3 & 60.8 & 66.0
& 25.2 & 67.9 & 56.4 & 61.6 \\
GPT-Terra
& 6.2 & 75.7 & 57.2 & \textbf{65.2}
& 0.0 & 54.9 & 59.2 & \textbf{57.0}
& 0.0 & 53.0 & 57.0 & \textbf{55.0}
& 0.0 & 73.2 & 64.2 & 68.4
& 0.9 & 68.6 & 62.4 & \textbf{65.4} \\
\midrule

iocextract
& 0.0 & 4.6 & 48.6 & 8.4
& 0.0 & 13.7 & 47.3 & 21.2
& 0.0 & 8.9 & 26.9 & 13.4
& 0.0 & 11.4 & 12.9 & 12.1
& 0.0 & 8.9 & 20.8 & 12.5 \\
URLExtract
& 0.0 & 22.0 & 48.6 & 30.3
& 0.0 & 42.1 & 44.7 & 43.3
& 0.0 & 15.2 & 27.3 & 19.5
& 0.0 & 26.2 & 12.6 & 17.0
& 0.0 & 24.9 & 20.3 & 22.4 \\
MSTICPy
& 0.0 & 1.3 & 47.6 & 2.5
& 0.0 & 40.4 & 48.0 & 43.9
& 0.0 & 22.8 & 26.2 & 24.4
& 0.0 & 5.9 & 12.6 & 8.0
& 0.0 & 4.4 & 20.5 & 7.2 \\

\bottomrule
\end{tabular}%
}

\label{tab:main_results}
\end{table*}

\subsubsection{Evaluation Metrics}
\label{sec:metrics}

We evaluate extraction performance on four IOC categories: URLs, domains, IP addresses, and filesystem artifacts, reporting precision, recall, and F1 for each.
For URLs, our primary metric uses query-insensitive matching: URLs are canonicalized to \texttt{scheme + host + path}, ignoring query strings, which avoids penalizing variations in query parameters that do not change the contacted endpoint.\footnote{We additionally report a stricter \emph{full-URL} metric that includes the query string in Appendix~\ref{apx:url_eval}.}
Under this canonicalization, the 2{,}890 raw IOC instances in Table~\ref{tab:ioc_label_distribution} correspond to 2{,}580 unique indicators used for evaluation.
Domains and IP addresses are scored by exact match after normalization, with excluded legitimate-service domains counted as false positives. Filesystem artifacts are evaluated within the labeling scope defined in \S\ref{sec:dataset-construction}.

We also define \textit{Response-failure rate} as the fraction of outputs that fail to produce a usable structured response, including unparsable, empty, prematurely terminated, or non-terminating outputs.
For precision, recall, and F1, such failures are treated as empty predictions.



\subsection{Findings}
\label{sec:findings}
This section analyzes LLM-based IOC extraction from multiple perspectives.
We begin with overall model performance against regex-based baselines, then analyze how performance varies with the recovery level of the target indicator and with input complexity.

\subsubsection{Benchmarking LLMs} 
\label{sec:eval}

\noindent\textbf{Frontier models lead, but robust IOC extraction remains challenging.} Table~\ref{tab:main_results} shows that robust IOC extraction from malicious scripts remains challenging for current LLMs.\footnote{For URL IOCs, the main results use the query-insensitive matching of \S\ref{sec:metrics}. The stricter URL matching results are reported separately in Appendix~\ref{apx:url_eval}.}
GPT-Terra and Gemini-Pro lead and are tied at around 65 F1.
Gemini-Flash follows, combining high precision with lower recall and a comparatively high response-failure rate. Among open-weight models, Qwen3.6-35B stands out, nearly matching the much larger Qwen3-C480B despite using far fewer active parameters. Scaling within the Qwen family otherwise yields clear gains, with the largest model achieving substantially higher recall and fewer response failures than the smaller variants. The 8B models remain well behind the larger systems, particularly on the more challenging script categories.

\noindent\textbf{Regex baselines remain competitive in recall, while LLMs achieve higher precision and recover beyond surface matching.}
The regex baselines provide a useful reference for surface extraction.
The strongest regex baseline, URLExtract, reaches only 22.4 F1, leaving a large gap to the leading LLMs.
However, several open-weight models show recall comparable to, or even below, the regex baselines, indicating that even surface-oriented IOC recovery is not consistently reliable.
$JS_{hp}$ shows a different pattern: despite low regex recall, even some 8B-scale models recover substantially more indicators.
This likely reflects that many samples follow regular downloader-style logic that iterates over candidate domains, making program-dependent recovery relatively tractable despite requiring more than literal string matching.

Even when recall gains are limited, open-weight LLMs generally achieve higher precision than the regex baselines.
This indicates that their advantage lies not only in recovering additional indicators, but also in being more selective and producing fewer unsupported predictions.
Appendix~\ref{apx:ioc_analysis} provides the corresponding per-IOC-type breakdown.

\noindent\textbf{Response failures vary across models.} 
Qwen3-8B usually fails before reaching the final answer, falling into repetitive reasoning loops such as repeated decoding explanations, sentinel values, or encoded substrings. Ministral3-8B, Gemma4-8B, Qwen3-C80B, and Gemini-Flash more often stop mid-generation after beginning a plausible analysis or deobfuscation. Thus, they have distinct error modes: failure to transition into the final answer, premature termination, or minor recoverable formatting errors.\footnote{We also tested larger token budgets, but the failed generations often stopped mid-generation in the same way, suggesting that these failures are not simply caused by insufficient output length.}

\subsubsection{Effect of IOC Recovery Level}
\label{sec:result_recovery_level}
We stratify each ground-truth IOC by its recovery level to examine how model performance changes as recovery requires progressively more reconstruction.
Table~\ref{tab:recall_by_level} reports recall by recovery level, with \lone{} further stratified by the number of recovery operations.

\textbf{\lzero{} IOCs are largely recoverable by stronger models.}
When indicators appear explicitly in the source, stronger models achieve high recall, with GPT-Terra reaching 91.7\% and Gemini-Pro 89.4\%. 
In contrast, smaller models show substantially lower recall even at this simplest level, indicating that the mere presence of an indicator in the input does not guarantee reliable extraction.

\begin{table}[t]
\centering
\footnotesize
\setlength{\tabcolsep}{3pt}
\renewcommand{\arraystretch}{1.05}
\caption{Micro recall (\%) by IOC recovery level. \lone{} stratified by recovery-operation count, with single-step cases split into reassembly (R) vs.\ transcode (T).}
\label{tab:recall_by_level}
\begin{tabular}{lcccccc}
\toprule
Model & \lzero{} & \multicolumn{4}{c}{Fixed-transf.} & Program-dep. \\
\cmidrule(lr){3-6}
      &          & 1\,(R) & 1\,(T) & 2 & $\ge 3$ &          \\
\textit{\# IOC instances} & (858) & (230) & (\phantom{0}85) & (156) & (\phantom{0}21) & (1,230) \\
\midrule
Qwen3-8B        & 66.9 & 27.8 & 16.5 & \phantom{0}2.6 & \phantom{0}0.0 & \phantom{0}1.5 \\
Ministral3-8B   & 78.7 & 33.0 & 16.5 & \phantom{0}3.8 & \phantom{0}0.0 & \phantom{0}2.8 \\
Qwen3.6-35B     & 80.2 & 59.1 & 37.6 & 16.0 & \phantom{0}0.0 & 14.7 \\
Qwen3-C480B     & 88.2 & 68.3 & 63.5 & 54.5 & \phantom{0}9.5 & 13.8 \\
Gemini-Flash    & 85.5 & 88.3 & 67.1 & 53.8 & 14.3 & 31.0 \\
Gemini-Pro      & 89.4 & 88.3 & 80.0 & 69.2 & \textbf{23.8} & \textbf{46.1} \\
GPT-Terra       & \textbf{91.7} & \textbf{90.9} & \textbf{82.4} & \textbf{69.9} & 19.0 & 35.4 \\
\bottomrule
\end{tabular}
\end{table}

\noindent\textbf{Recovery degrades rapidly as reconstruction becomes deeper.}
For \lone{} IOCs, recall declines steadily as recovery requires more operations, and a single transcode (T) is harder than a single reassembly (R). GPT-Terra, for example, falls from 90.9\% (single reassembly) and 82.4\% (single transcode) to 69.9\% at two operations and 19.0\% at three or more.
Multi-operation chains are particularly challenging because they require the model to accurately invert multiple transformations in sequence.
Smaller models start from substantially lower recall and deteriorate even more sharply. Thus, even when the required transformations are deterministic and known in principle, reliably composing them remains difficult for current LLMs.

\noindent\textbf{\ltwo{} recovery remains difficult even for frontier models.}
When recovery depends on sample-specific program logic, recall drops sharply across all models. Even the frontier models recover only 31--46\% of these IOCs, and all other models perform substantially worse.
Notably, Qwen3-C480B performs competitively on simpler recovery levels but reaches only 13.8\% recall at this level.
This suggests that program-dependent recovery remains a major limitation of current models, while also exposing the clearest differences in their ability to reason over program-specific logic.
The recovery-level trend remains consistent within each dataset (Appendix Table~\ref{tab:recall_by_level_per_dataset}).

\noindent\textbf{Frontier models struggle with compositional and iterative value recovery.}
To better understand where the frontier models succeed and fail, we break down their recall by recovery operation in Appendix~\ref{apx:transform-recall}. 
For \textbf{\lone}, models generally handle many isolated operations well, including literal concatenation, split/join, and some single-step transcodes, but recall drops when reconstruction requires multiple operations (Table~\ref{tab:transform_recall}).
The models also exhibit different strengths: the Gemini models are substantially stronger than GPT-Terra on constant propagation, whereas GPT-Terra performs better on Base64-to-UTF-16LE decoding.
The \textbf{\ltwo{}} results further show that difficulty rises when value recovery depends on iterative program execution: loop-constructed values stand out as the hardest mechanism for all three models (Table~\ref{tab:l2_mechanism_recall}). 
Taken together, these results suggest that the main bottleneck is composing multiple transformations and tracking values through iterative execution.

\begin{figure}[t]
    \centering
    \includegraphics[width=0.9\linewidth]{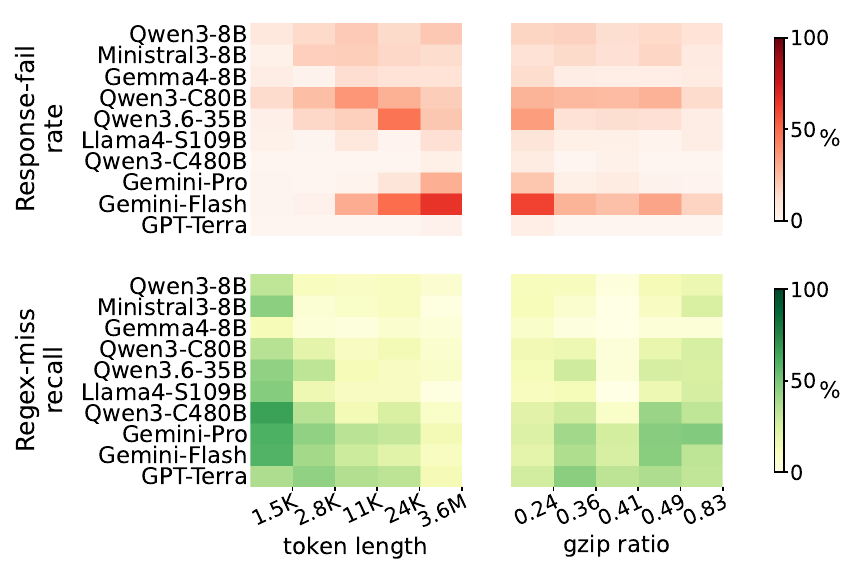}
    \caption{Effect of input complexity on LLM-based IOC extraction. Columns are binned by token length or gzip compression ratio. The top row reports Response-failure rate, where darker is worse; the bottom row reports Regex-missed Recall, where darker is better.}
    \label{fig:complexity_heatmap}
\end{figure}

\begin{table*}[t]
\centering
\footnotesize

\setlength{\tabcolsep}{2.8pt}
\renewcommand{\arraystretch}{0.85}
\caption{
False-positive taxonomy, grouped by provenance; the assignment precedence is deterministic (Appendix~\ref{apx:fp_taxonomy}).
\textbf{Disposition} (operational handling if the error survives triage):
\costbox{cO0}~\emph{Filterable},
\costbox{cO1}~\emph{Recoverable},
\costbox{cO2}~\emph{Review-inducing},
\costbox{cO3}~\emph{Misleading}.
\gold{G}/\pred{P} denote the reference IOC and prediction.
}
\begin{NiceTabular}{
    c
    >{\centering\arraybackslash\scriptsize}m{0.03\textwidth}
    >{\raggedright\arraybackslash}m{0.14\textwidth}
    >{\raggedright\arraybackslash}m{0.45\textwidth}
    >{\raggedright\arraybackslash\scriptsize}m{0.23\textwidth}
    >{\centering\arraybackslash}m{0.04\textwidth}
}
\toprule

\textbf{Group}
& \textbf{}
& \textbf{Category}
& \textbf{Classification criterion}
& \textbf{Example}
& \textbf{Disp.}
\\

\midrule

\Block{6-1}{\grouplabel{Sample}{-traceable}}
& \Block{4-1}{\emph{IOC}}
& Type mismatch
& The normalized value equals a reference IOC of a \emph{different} type.
& $G_{ip}$: \gold{1.2.3.4}\quad $P_{dom}$: \pred{1.2.3.4}
& \costbox{cO1}
\\
\cmidrule(lr){3-6}
&
& Normalization error
& Malformed, yet \emph{deterministically} normalizes to a reference IOC.
& G: \gold{https://a.co}\quad P: \pred{htps:/a.co}
& \costbox{cO1}
\\
\cmidrule(lr){3-6}
&
& Granularity
& The right entity at a coarser or finer level.
& G: \gold{a.b.com/x/y}\quad P: \pred{a.b.com/x}
& \costbox{cO1}
\\
\cmidrule(lr){3-6}
&
& Near-copy
& A distorted copy of a reference IOC: a mis-cut boundary or altered characters.
& G: \gold{24sports.ca}\quad P: \pred{sports.ca}
& \costbox{cO3}
\\

\cmidrule(l){2-6}

& \Block{2-1}{\emph{non-}\\\emph{IOC}}
& Benign shared service
& A benign shared platform in the sample, tied to no reference indicator.
& P: \pred{ifconfig.me/ip}
& \costbox{cO0}
\\
\cmidrule(lr){3-6}
&
& Source non-IOC
& A source string that is neither an indicator nor a benign service.
& P: \pred{fileserver}
& \costbox{cO2}
\\

\midrule

\Block{4-1}{\grouplabel{Sample}{-untraceable}}
& \Block{4-1}{}
& Prompt echo
& A value copied from the instruction template.
& P: \pred{cdn.example.test}
& \costbox{cO0}
\\
\cmidrule(lr){3-6}
&
& Generic placeholder
& A source-absent reserved/private/example pattern, or a hallucinated benign service.
& P: \pred{192.168.1.1}
& \costbox{cO0}
\\
\cmidrule(lr){3-6}
&
& Ungrounded invalid
& Syntactically invalid and not traceable to any reference IOC or source.
& P: \pred{maa1}
& \costbox{cO0}
\\
\cmidrule(lr){3-6}
&
& Unattributed
& A specific, valid-looking value with no trace in source, prompt, priors, or gold.
& P: \pred{downprofits.ru}
& \costbox{cO2}\kern-0.35ex\costbox{cO3}
\\

\bottomrule
\end{NiceTabular}

\label{tab:fp_taxonomy_v7}

\end{table*}

\subsubsection{Effect of Input Complexity}
\label{sec:result_code_complexity}
We analyze LLM extraction performance along the two dimensions introduced in \S\ref{sec:dataset_complexity}: input size (token length) and code compressibility (gzip compression ratio; a lower ratio indicates more redundant content).
We use response-failure rate to measure response reliability and \textit{Regex-missed Recall} to capture recovery capability beyond surface-oriented regex extraction, with the latter computed over ground-truth indicators missed by the regex baseline.

\noindent\textbf{Longer inputs degrade both recovery capability and response reliability.}
Response-failure rates tend to increase with token length, although the pattern varies across models.
Regex-missed Recall also generally declines as token length increases, indicating that longer inputs make recovery beyond surface matching increasingly difficult.
Notably, Qwen3-C480B maintains low response-failure rates across most bins while its Regex-missed Recall still declines with length.
Thus, reliably completing a structured response does not necessarily imply robust IOC recovery from long inputs.

\noindent\textbf{Code complexity shows a weaker and different effect.}
Lower gzip-ratio bins show a weak tendency toward higher response-failure rates, suggesting that highly redundant inputs may reduce output reliability.
In contrast, Regex-missed Recall tends to be somewhat higher in higher gzip-ratio bins, although the pattern varies across models.

\begin{figure*}[t]
    \centering
    \includegraphics[width=1\linewidth]{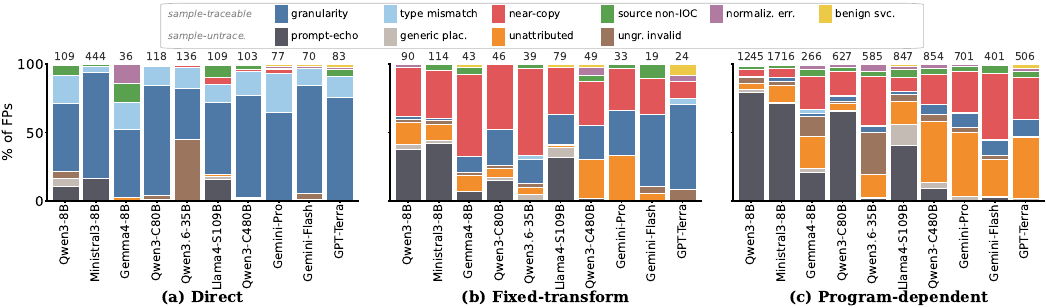}
    \caption{Per-model false-positive composition within each IOC recovery level. Numbers above bars indicate total false positives.}
    \label{fig:fp_by_level}
\end{figure*}

\subsection{Failure Modes Analysis}
\label{sec:fp_analysis}

\subsubsection{False Positive Category}

LLMs can generate plausible-looking IOCs that are not directly supported by the input, and such errors can reduce analyst trust and increase downstream verification cost. 
Therefore, a fine-grained analysis of false positives is necessary for understanding where LLMs are vulnerable as IOC extractors and what types of mitigation are needed.

To support systematic error analysis, we define a taxonomy of false positives, summarized in Table~\ref{tab:fp_taxonomy_v7}.
Its top-level split is \emph{provenance}, based on whether a prediction is traceable to the sample exactly or through a bounded normalization or near-match relation. Traceable errors are further divided into distorted reference IOCs and non-reference source strings. Each category is also associated with its likely operational disposition.
A deterministic precedence rule assigns every false positive to exactly one of the ten categories; full rules are given in Appendix~\ref{apx:fp_taxonomy}.

\subsubsection{Results}
\label{sec:fp_results}
Figure~\ref{fig:fp_by_level} shows each model's false-positive composition within each IOC recovery level\footnote{Because a false positive has no intrinsic recovery level, we associate each error with the recovery difficulty of the reference IOC it most plausibly replaces or accompanies.}, revealing clear differences in model error profiles.

\noindent\textbf{Smaller models fall back to prompt-echo errors at higher recovery levels.}
Several smaller open-weight models exhibit a distinctive failure mode: when they fail to recover an IOC from the script, they frequently emit IOC examples copied from the prompt.
Prompt echo dominates the false positives of Qwen3-8B and Ministral3-8B, and is disproportionately associated with missed \ltwo{} IOCs rather than directly observable ones.
This suggests a shortcut-like fallback: when recovery from the script becomes difficult, smaller models often reuse readily available IOC examples from the prompt instead of analyzing the malware sample.

\noindent\textbf{Prompt-echo errors decrease with scale within a model family.} The same pattern persists in some larger open-weight models but decreases sharply with scale. Within the Qwen3-Coder family, Prompt echo accounts for 53\% of false positives from Qwen3-C80B but only about 8\% from Qwen3-C480B. The remaining errors of the larger model instead shift toward categories such as Granularity and Unattributed. This within-family comparison suggests that scaling substantially reduces reliance on prompt-derived outputs.


\noindent\textbf{Frontier-model errors shift from Prompt echo to reconstruction failures.}
At the frontier end, false positives are less dominated by prompt-derived artifacts and instead reflect failures during reconstruction.
Many remain grounded in the sample, appearing as Near-copy or Granularity errors in which the predicted IOC is closely related to the underlying evidence but does not match the exact indicator.
This pattern is consistent with known limitations of LLMs in character-level string manipulation~\cite{shin2024large,wang2024stringllm}. As recovery becomes more difficult, Unattributed errors also become more prominent: these predictions are no longer directly traceable to the source, suggesting that reconstruction can drift beyond the available evidence during deobfuscation. Thus, frontier models fail less often by reverting to prompt examples, but still struggle with both precise reconstruction and maintaining grounding throughout recovery.

Taken together, recovery level reveals a qualitative capability gap in how models fail: smaller models increasingly lose sample grounding when script-specific reconstruction is required, whereas frontier models tend to remain closer to the malware evidence. Increased model capacity shifts errors in the latter direction; in \S\ref{sec:adaptation} we show that adaptation and tool augmentation reproduce the same shift in a small model without scaling. Operationally, stronger models therefore produce fewer obviously spurious indicators, but their residual errors are more plausible and may require closer analyst verification.
\section{Improving Small-Model IOC Recovery with Tools and Adaptation}
\label{sec:adaptation}
The benchmark results expose two prominent weaknesses of 8B-scale models: they perform reasonably on directly observable indicators but struggle once IOC recovery requires reconstruction (\S\ref{sec:result_recovery_level}), and they lose grounding under difficult cases, as reflected in frequent prompt-echo false positives (\S\ref{sec:fp_analysis}).
We therefore examine two targeted approaches: 1) deterministic string utilities, which offload exact decoding and reconstruction, and 2) task-specific adaptation, which teaches the model the analysis behavior required for IOC extraction. We use \benchmark{} as a testbed to evaluate these approaches separately and together in a controlled case study on a single open-weight model.

\subsection{Study Design}
\label{sec:case-study-design}
We conduct the case study on Qwen3-8B, the strongest 8B-scale model in our evaluation (\S\ref{sec:eval}), and compare four variants under the shared inference setup of \S\ref{sec:inference-setup}: the base model, the base model with tool access, the task-adapted model, and the task-adapted model with tool access.

\noindent\textbf{String-utility tool augmentation.}
We provide deterministic utilities for common decoding, string reconstruction, and reversible transformations, including operations such as Base64 decoding, concatenation, split/join, reversal, and XOR. The utilities operate only on strings and integer arrays and do not execute code, access the filesystem, or open network connections, preserving the static-analysis setting. Tool outputs are returned to the model for subsequent reasoning; the full tool set and agent configuration are given in Appendices~\ref{apx:prompt} and~\ref{apx:inference_config}.

\noindent\textbf{Task-specific adaptation.}
We adapt Qwen3-8B using supervised fine-tuning followed by Group Relative Policy Optimization (GRPO)~\cite{shao2024deepseekmath} with a task-specific verifiable reward. Training uses an additional 1{,}100 JavaScript malware samples from 2015--2016, drawn from the same collection as $JS_{hp}$ but disjoint from the evaluation set, together with benign JavaScript samples. The objective encourages valid structured outputs, correct verdicts, and precise IOC extraction. Importantly, the adaptation contains no supervision for tool use, allowing us to evaluate whether adaptation and tool augmentation provide complementary gains. Full training and reward details are provided in Appendix~\ref{apx:training}.

\noindent
\hangindent=1.2em
\hangafter=1
\textbullet\hspace{0.5em}\textbf{Supervised fine-tuning.}
For SFT, a strong teacher model generates reasoning-rich responses with intermediate analysis and structured IOC outputs, and the student is trained on the filtered synthetic supervision using the standard language-modeling objective. 

\noindent
\hangindent=1.2em
\hangafter=1
\textbullet\hspace{0.5em}\textbf{Reward-based optimization.}
Starting from the SFT model, we apply GRPO with a task-specific verifiable reward designed for structured IOC extraction.
The reward jointly encourages accurate IOC extraction through separate precision and recall terms, rewarding coverage of supported indicators while discouraging unsupported over-prediction.
It also includes auxiliary terms for output formatting, malicious/benign classification, and consistency between declared IOC status and emitted indicators.
We also introduce longer and more complex scripts gradually during training for optimization stability. Full reward definitions and training details are provided in Appendix~\ref{apx:training}.



\begin{figure}[t]
    \centering
    \includegraphics[width=1\linewidth]{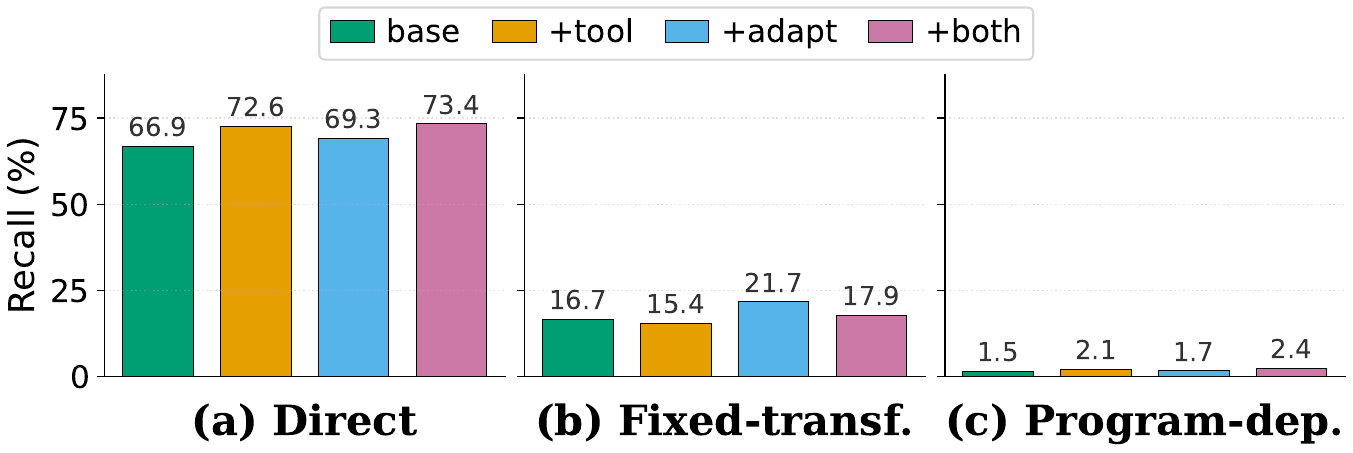}
    \caption{Recall by IOC recovery level for the variants.}
    \label{fig:qwen_recall_level}
\end{figure}
\begin{figure}[t]
    \centering
    \includegraphics[width=1\linewidth]{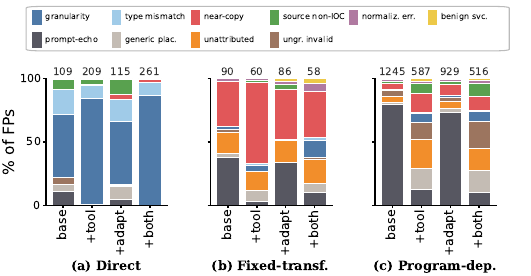}
    \caption{False-positive composition for the variants.}
    \label{fig:qwen_fp_comp}
\end{figure}

\subsection{Results}
\label{sec:case-study-results}
Figures~\ref{fig:qwen_recall_level} and~\ref{fig:qwen_fp_comp} show how tool access and adaptation affect both recovery capability and false-positive behavior; Appendix Figure~\ref{fig:qwen3_variants} provides the overall performance across script categories.

\noindent\textbf{Tool access improves grounding.}
String-utility access changes both recall and the character of the model’s errors. Its largest recall gain appears on \lzero{} indicators, increasing from 67\% to 73\%; in contrast, recall decreases on \lone{} and increases only slightly on \ltwo{}.
Tool use also introduces additional tool interactions and intermediate reasoning, increasing generation overhead and causing more parsing failures in some cases, which contributes to the reduced \lone{} recall.
At the same time, tool access sharply suppresses Prompt echo, which dominates the base model’s false positives. 
Together, these results suggest that tool access primarily improves grounding and helps the model recover indicators that are directly present in the script.

\noindent\textbf{Adaptation helps most with \lone{}.}
Adaptation exhibits a different recovery profile.
Its largest gain appears on \lone{} indicators, where recall increases from 17\% to 22\%, while the improvement on \lzero{} indicators is more modest.
Within \lone{}, adaptation improves single-step reassembly recall from 28\% to 38\%, while single-step transcode recall changes little (16\% to 19\%).
This suggests that adaptation helps with recurring reassembly patterns, while reliable transcoding remains challenging.
Adaptation also reduces prompt-echo errors, although less sharply than tool access.

\noindent\textbf{Improvements do not extend to \ltwo{} recovery.}
Despite clear gains in grounding and \lone{} recall, \ltwo{} recall remains at only 1--3\% across all four variants.
These approaches therefore improve grounding and fixed reconstruction without materially extending recovery through sample-specific program logic.
This remaining bottleneck is not specific to Qwen3-8B; \ltwo{} recovery is the hardest part even for the frontier models in our benchmark.

\noindent\textbf{Both approaches improve precision, with the combined variant performing best.}
Appendix Figure~\ref{fig:qwen3_variants} shows that both tool access and adaptation substantially improve precision, with the combined model achieving the highest precision, raising it from 31\% for the base variant to 48\%. These gains indicate that these approaches not only recover more IOCs but also reduce unsupported predictions.

\noindent\textbf{Scope.}
This case study is intended to demonstrate that the failure modes \benchmark{} surfaces are actionable, rather than to establish generality across models; broader studies, including tool-aware adaptation, are discussed in \S\ref{sec:discussion}.

\section{Discussion}
\label{sec:discussion}

\noindent \textbf{Implications for Analyst Workflows.}
Our results show that current LLMs are not yet reliable enough for standalone IOC extraction. Even the strongest model reaches only 65.4 F1, and recall drops sharply for Program-dependent indicators. Moreover, models exhibit different failure modes, suggesting that their outputs require verification tailored to their error profiles.
At the same time, our controlled study on a small open-weight model shows that deterministic string utilities and task-specific adaptation can improve grounding and precision. 
These findings motivate a layered static-analysis workflow in which conventional extractors, LLMs, and deterministic utilities support initial IOC recovery, while unresolved cases are escalated to specialized program analysis or analyst review.

\noindent \textbf{Ground-Truth Scope.}
In this study, we focus on URLs, domains, IP addresses, and filesystem artifacts because they collectively capture both network infrastructure and host-side traces and commonly arise in script-based malware.
Other IOC types remain outside the current scope.
For example, registry artifacts may require a different representation, as the threat-relevant unit can involve not only a key path but also the associated value name, value data, and registry operation.
Including them would therefore require additional annotation and matching rules beyond those considered here.
Such artifact types remain natural targets for future extensions of the benchmark.



\noindent\textbf{Data Contamination}.
Since the Hynek Petrak collection (2015--2017) is publicly available, pretraining contamination on $JS_{hp}$ is a potential concern.
In contrast, the more recent $JS_f$, $PS_f$, and $VBS_f$ datasets were collected after the documented knowledge cutoffs of all evaluated models except GPT-Terra, and therefore cannot overlap with their pretraining data under those cutoffs. 
The main findings persist across both the historical and recent datasets. GPT-Terra is the only exception for which the training window may partially overlap with the recent data, so contamination cannot be ruled out for this model. Nevertheless, its performance remains far from saturated on the recent subsets, reaching only 65.2, 57.0, and 55.0 F1 on $JS_f$, $PS_f$, and $VBS_f$, respectively. These results do not eliminate contamination as a possibility for GPT-Terra, but they provide no indication that its performance is trivially explained by memorization.

\vspace{2pt} \noindent \textbf{Future Work.}
Future work can extend \benchmark{} to additional script languages and IOC types. 
Our tool-use study can also be expanded to multiple models evaluated with the deterministic utilities, allowing controlled comparison across model families. 
These experiments could further motivate tool-aware adaptation, where models learn when to invoke external utilities and how to integrate their outputs into IOC predictions. 
Finally, we plan to maintain the benchmark and leaderboard through periodic updates covering new models and emerging malware.


\section{Conclusion}
We presented \benchmark{}, a benchmark for evaluating whether LLMs can recover concrete IOCs directly from malicious scripts under static analysis. 
\benchmark{} targets artifact-level recovery of URLs, domains, IP addresses, and filesystem artifacts from JavaScript, PowerShell, and VBScript malware, with manually verified labels curated for static recoverability.  
Our results show that static IOC extraction remains challenging even for frontier models; they perform well on directly exposed IOCs but struggle with deeper reconstruction.
A fine-grained false-positive analysis shows that their error profiles vary across model capabilities. Tool augmentation and task-specific adaptation improve grounding and precision, but Program-dependent recovery remains a key challenge. Overall, \benchmark{} provides a systematic testbed for measuring progress toward reliable LLM-assisted IOC extraction.



\clearpage
\bibliographystyle{plain}
\bibliography{reference}
\clearpage
\appendix
\section{Leaderboard}
\label{apx:leaderboard}
\begin{figure}[t]
    \centering
    \includegraphics[width=1\linewidth]{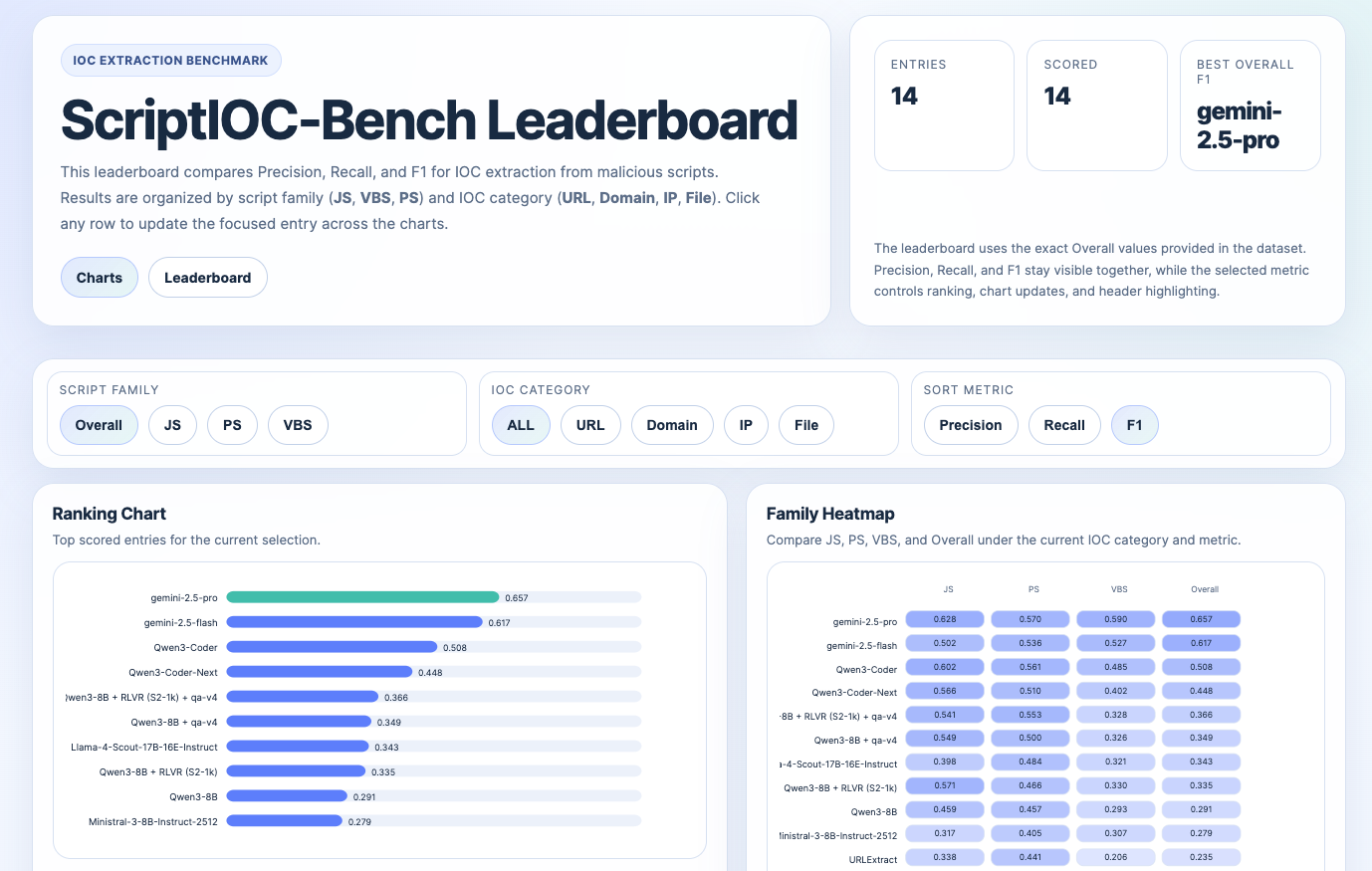}
    \caption{Screenshot of \benchmark{} leaderboard for tracking static IOC extraction performance across LLMs.}
        \label{fig:leaderboard}
\end{figure}

We will release our \benchmark{} as a public leaderboard (Figure~\ref{fig:leaderboard}) to encourage reproducible evaluation and community participation.
The leaderboard provides a shared testbed for evaluating LLM-based static IOC extraction systems on realistic malicious scripts.
By comparing open-weight LLMs, frontier models, tool-augmented agents, and conventional rule-based baselines under the same protocol, the leaderboard makes it possible to track progress and expose remaining failure modes.
Through broader participation, we hope to accelerate the development of reliable LLM-assisted IOC extraction methods and contribute to practical threat-intelligence automation.

\section{Regex Baseline Details}
\label{apx:regex_baselines}
URL candidates are truncated at the first quote, whitespace, or control character, and are discarded if they exceed 500 characters or lack a valid \texttt{http}, \texttt{https}, or \texttt{ftp} scheme and host. 
Domain candidates must have a 2--6 character alphabetic TLD; we discard candidates whose TLD is a known file extension, whose labels contain common JavaScript keywords or \texttt{\_0x...} obfuscation tokens, or whose labels are single-character, purely numeric, or version-like. 
For file candidates, we remove paths shorter than five characters and, for MSTICPy path matches, additionally filter strings containing \texttt{\_0x[0-9a-f]+}, \texttt{function(}, or \texttt{return ;} to avoid JavaScript fragments being misclassified as Linux paths.

\section{Training Details}
\label{apx:training}
We adapt Qwen3-8B through a two-stage training procedure. The first stage performs supervised fine-tuning (SFT), and the second stage continues from the supervised fine-tuned checkpoint using reinforcement learning with verifiable rewards. 

Both stages train LoRA adapters (rank 128, $\alpha{=}32$, dropout 0.05, applied to all attention and MLP projections) over the open-weight Qwen3-8B base model on three NVIDIA H200 GPUs in bfloat16. The merged checkpoint is evaluated with the same vLLM map--reduce pipeline as the other Qwen3-8B rows (three repetitions, deterministic decoding at $T{=}0$, 32,768-token served context, 8,192-token generation budget).

\noindent\textbf{SFT.}
The first stage trains Qwen3-8B on the IOC extraction prompt--completion pairs distilled from Gemini 2.5 Pro. The source data includes JavaScript malware samples from multiple difficulty levels, as well as benign JavaScript samples. After filtering to fit the training context, the SFT
dataset contains 1,823 prompt--completion pairs.
Training uses AdamW (learning rate $1\times10^{-5}$, linear schedule, warmup ratio 0.03) with an effective batch of 48 prompts, capped at 1,000 steps; the checkpoint for the next stage is selected by training loss.

\noindent\textbf{Silver IOC curation.}
Report-derived IOC labels serve as the gold reference set $\mathcal{G}(x)$. Such reports may include dynamically observed indicators and miss IOCs that are statically recoverable from the script, and manually auditing every training sample is impractical at training scale. We therefore augment $\mathcal{G}(x)$ with a silver set $\mathcal{S}(x)$ of additional teacher-extracted candidates that pass syntactic validity checks and are not already present in $\mathcal{G}(x)$. The combined references $\mathcal{G}(x) \cup \mathcal{S}(x)$ are used by the IOC reward terms in the subsequent GRPO stage (\S\ref{apx:reward}), with lower confidence assigned to silver-only matches.

\noindent\textbf{GRPO.}
The second stage continues from the SFT checkpoint using the clipped GRPO objective ($\epsilon{=}0.2$) with group-relative reward normalization, AdamW (learning rate $1\times10^{-5}$, warmup ratio 0.05), and an effective batch of 288 prompts with 8 rollouts each, generated with vLLM at $T{=}0.7$ under a 4,096-token completion cap. Training follows a two-stage length curriculum: 500 steps on prompts of at most 4,096 tokens, then 500 further steps on prompts of at most 16K tokens.

\subsection{Reward Function}
\label{apx:reward}

This subsection details the GRPO reward introduced above. Unlike single-label classification, IOC extraction requires both a malicious/benign verdict and a typed set of indicators, and outputs are typically only partially correct. The reward is therefore tailored to this task structure, comprising five terms that each address a distinct failure mode of malicious-code analysis:

\begin{align}
R(y; x) =\; & \lambda_f R_{\text{fmt}}(y) + \lambda_c R_{\text{cls}}(y; x) + \lambda_s R_{\text{status}}(y) \nonumber \\
& + \lambda_p R^{\text{prec}}_{\text{ioc}}(y; \mathcal{G}(x), \mathcal{S}(x)) + \lambda_r R^{\text{rec}}_{\text{ioc}}(y; \mathcal{G}(x), \mathcal{S}(x)),
\end{align}

where $\mathcal{G}(x)$ and $\mathcal{S}(x)$ denote the gold and silver reference sets defined above (\textbf{Silver IOC curation}).

\noindent\textbf{Format and classification.}
$R_{\text{fmt}} \in \{0, 0.5, 1\}$ rewards parseable outputs, with full 
credit when the per-type IOC status block is present. $R_{\text{cls}}$ is 
$1$ if the predicted malicious/benign verdict matches the gold label and 
$0$ otherwise.

\noindent\textbf{Self-consistency.}
$R_{\text{status}}$ rewards agreement between the declared per-type IOC 
status (\texttt{present}/\texttt{suspected}/\texttt{absent}) and the IOCs 
actually emitted. A mismatch occurs when the model declares \texttt{absent} 
but emits IOCs, or declares \texttt{present} but emits none.

\noindent\textbf{IOC precision and recall.}
Predictions are matched against gold $\mathcal{G}(x)$ and silver 
$\mathcal{S}(x)$ reference sets, with credit weighted by reference 
reliability: $1.0$ for IOCs in $\mathcal{G}(x) \cap \mathcal{S}(x)$, 
$0.85$ for gold-only, and $0.40$ for silver-only. Per-type precision and 
recall are computed for $\{\texttt{host}, \texttt{ip}, \texttt{file}, 
\texttt{url}\}$ and averaged with equal weights.

To prevent reward hacking, we apply two countermeasures. 
\emph{(i) Precision penalty:} if a sample has only false positives, 
precision becomes negative (down to $-0.45$), with stronger penalty when 
no gold IOCs exist (penalizing fabrication) than when gold IOCs are 
simply missed. \emph{(ii) Recall dampening:} when the model emits both 
matched and unmatched IOCs, recall is divided by $1 + 0.65 \cdot E$ 
where $E$ is the number of false positives, preventing the model from 
maximizing recall through mass enumeration.


\noindent\textbf{Reward weights.}
We use $\lambda_f{=}0.5$, $\lambda_c{=}1.0$, $\lambda_s{=}2.0$, and 
$\lambda_p{=}\lambda_r{=}4.0$, emphasizing IOC quality over format and 
verdict alone.

\subsection{Hyperparameters}
\label{app:reward-hparams}

Table~\ref{tab:reward-hparams} summarizes the reward weights and internal 
constants used throughout training.
Type weights for $\{\texttt{host}, \texttt{ip}, \texttt{file}, \texttt{url}\}$ are equal; \texttt{file} subcomponent weights are $0.60$/$0.25$/$0.15$ for file names/dir roots/subdirs.

\begin{table}[h]
\centering
\footnotesize
\caption{Reward hyperparameters.}
\begin{tabular}{lll}
\toprule
\textbf{Symbol} & \textbf{Value} & \textbf{Description} \\
\midrule
$\lambda_f$ & 0.5 & Format weight \\
$\lambda_c$ & 1.0 & Classification weight \\
$\lambda_s$ & 2.0 & Self-consistency weight \\
$\lambda_p$ & 4.0 & IOC precision weight \\
$\lambda_r$ & 4.0 & IOC recall weight \\
\midrule
$w_{\text{both}}$ & 1.00 & Gold $\cap$ silver match weight \\
$w_{\text{gold}}$ & 0.85 & Gold-only match weight \\
$w_{\text{silver}}$ & 0.40 & Silver-only match weight \\
$\alpha$ & 1.2 & Per-type FP coefficient in precision \\
$\beta$ & 0.45 & Aggregate FP penalty magnitude \\
$\gamma$ & 0.65 & Recall dampening rate \\
$\rho_{\text{sus}}$ & 0.10 & Suspected-status partial recall credit \\
\bottomrule
\end{tabular}
\label{tab:reward-hparams}
\end{table}

\section{Pipeline Prompts}
\label{apx:prompt}
\noindent\emph{Note: the boxes below are simplified excerpts. The full
prompts (verbatim) are released with the evaluation code.}

\smallskip
We use a \textbf{single-pass} prompt for short samples and a
\textbf{map}/\textbf{reduce} pair for long samples. Each has a base
variant and a tool-augmented variant; the latter only appends a
\textsc{Tool Usage} block to the base system prompt.

\begin{promptbox}[Single-pass extraction --- system prompt]
\scriptsize
Interpret the script, deobfuscate as needed, classify it as
\emph{malicious} or \emph{benign}, and extract deterministic IoCs.
Output three sections:

\begin{verbatim}
<analysis>      reasoning, classification justification
<deobfuscated_code>  minimal behaviour-focused reconstruction
<answer>
{ "classification": "malicious" | "benign",
  "ioc_status": {host, ip, file, url -> P|A|S},
  "url_groups": [{canonical_url, placeholder_values}],
  "host":[...], "ip":[...], "file":[...], "url":[...] }
\end{verbatim}

\textbf{Status (P/A/S):} \emph{present} (concrete value recovered ---
list it), \emph{absent} (no evidence), \emph{suspected} (strong
evidence, no concrete value after attempted decoding).

\textbf{Rules.} Recovery before \emph{suspected}; only values
explicitly produced by the code.
\end{promptbox}

\begin{promptbox}[Map step --- system prompt]
\scriptsize
Analyse \emph{one chunk} of a long script. Record partial observations
only; do not classify; do not emit \texttt{<analysis>} /
\texttt{<deobfuscated\_code>} / \texttt{<answer>}.

\begin{verbatim}
<chunk_summary>      one paragraph
<deobfuscation_notes>  decodings from this chunk;
  if a blob (base64/hex/char-codes/reversed)
  cannot be fully decoded here, paste it
  verbatim with its suspected encoding.
<candidate_iocs>     {host, ip, file, url}
                     (concrete values only)
<unresolved>         deps on other chunks
\end{verbatim}
\end{promptbox}

\begin{promptbox}[Reduce step --- system prompt]
\scriptsize
Same as the single-pass prompt (same three sections, same schema, same
status / recovery rules), plus:

\smallskip
\textbf{Cross-chunk synthesis.} When a finding is \texttt{<unresolved>}
in one chunk but another supplies the missing key/fragment, complete
the decoding from the combined evidence and cite the recovered value.
Never invent values the evidence does not support.
\end{promptbox}

\begin{promptbox}[Appended block --- \textsc{Tool Usage} (tool variants)]
\scriptsize
Call the appropriate tool rather than guessing; guessed decodings are
not acceptable.

\begin{verbatim}
"TVqQ..."        -> base64_decode
"\x48..", hex     -> hex_decode
Chr(N)& / fromCharCode -> char_codes_to_string
"%2F.."          -> url_decode
xor + key         -> xor_hex
StrReverse("...") -> string_reverse
"pow"&"ersh"      -> concat_strings
"p-o-w".Replace   -> split_and_join
"[.]" -> "."     -> string_replace
ROT-N             -> rot_n
\end{verbatim}

Call tools \emph{before} writing the tagged sections; cite the exact
tool output. 
Mark \emph{suspected} only after a tool fails or returns
non-text. 
\emph{Map} variant: list cross-chunk dependencies in
\texttt{<unresolved>} instead of guessing.
\emph{Reduce} variant:
tools are used only to finish blobs left \texttt{<unresolved>} by
the map step.
\end{promptbox}

\section{Model and inference configurations}
\label{apx:inference_config}
We evaluate three model groups. The \textbf{8B-scale open-weight} group includes \textit{Qwen3-8B}, \textit{Ministral-3-8B-Instruct-2512}, and \textit{Gemma-4-E4B-it}, served locally with each model's native context length as \texttt{max\_model\_len}. The \textbf{larger open-weight} group includes \textit{Qwen3-Coder-Next}, \textit{Qwen3-Coder-480B-A35B-Instruct}, \textit{Qwen3.6-35B-A3B}, and \textit{Llama-4-Scout-17B-16E-Instruct}, served locally or via the HuggingFace Inference Router under long-context configurations (\textit{Qwen3.6-35B-A3B} at a $262$K served context, evaluated with three repetitions at $T=0$). The \textbf{frontier proprietary} baselines are \textit{Gemini 2.5 Pro} and \textit{Gemini 2.5 Flash}, accessed through Google's native SDK with \texttt{max\_model\_len=1{,}048{,}576}, \texttt{max\_output\_tokens=32{,}768}, and automatic thinking-budget configuration. \textit{GPT-5.6-terra} (\texttt{gpt-5.6-terra}), accessed via its API with a $922$K served context and evaluated with reasoning enabled; temperature was not configurable under this setting, so we used the API default sampling configuration.
 
Per-model output budgets are sized to fit within the served context window: 8{,}192 tokens for Qwen3-8B, 4{,}096 for its agent variants (leaving room for tool-call history within the same served context), 32{,}768 for Ministral3-8B, 16{,}384 for Gemma4-8B, and 32{,}768 for the Gemini models.
 
\noindent\textbf{String-utility tool set.}
The string utilities exposed to the agent cover three categories of deterministic operations: \emph{decode} tools convert encoded representations into plaintext, including Base64, hexadecimal, URL encoding, and numeric character-code arrays; \emph{recover} tools reconstruct obfuscated strings from script-level operations such as concatenation, split-and-join patterns, substring replacement, and string reversal; and \emph{transform} tools apply reversible string or byte transformations, including ROT-$n$, XOR, Base64 encoding, and hexadecimal encoding. All utilities are pure functions over strings and integer arrays: they do not execute code, access the filesystem, or open network connections, keeping the setting strictly static.

For agent settings, we wrap the same vLLM-served model with a Qwen-Agent tool-calling wrapper that exposes these utilities and feeds tool outputs back into subsequent model calls. We enforce a per-sample limit of ten model calls.

\section{Map--reduce Long-context Handling}
\label{apx:chunking}

Samples exceeding the per-chunk code budget are processed with map--reduce. Each chunk produces partial findings, which are aggregated by a final reduce call into the structured output. Map outputs are truncated to 1{,}500 tokens, so a sample split into $N$ chunks contributes approximately $N\cdot1{,}500+1{,}000$ tokens to the reduce prompt.

Let $L_{\text{model}}$ be the served context length, $P_{\text{prompt}}\approx1{,}000$ the prompt overhead, and $T_{\text{out}}$ the reserved generation budget. We set the per-chunk code budget to
\[
B \;=\; \max\!\left(1024,\;\; 0.85 \cdot \bigl(L_{\text{model}} - (P_{\text{prompt}} + T_{\text{out}}) - 2048\bigr)\right).
\]
where the margin accounts for tokenizer drift and decoded-length expansion.\footnote{For 8 of 634 extreme-length samples, all evaluated with Qwen3-8B and split into more than 100 chunks, the accumulated map summaries reach the served context limit at the reduce stage; these cases are counted as response failures. All other models remain within the reduce budget.}

\section{URL extraction performance}
\label{apx:url_eval}
Table~\ref{tab:url_strictness} reports URL extraction performance under three matching criteria: endpoint-level matching, endpoint-plus-query-key matching, and full-URL matching.
Overall, F1 scores decrease as the matching criterion becomes stricter, indicating that models more often recover the correct endpoint than the exact query structure or full URL string.
\begin{table}[t]
\caption{URL extraction performance by matching strictness, micro-averaged over IOC instances.}
\centering
\footnotesize
\setlength{\tabcolsep}{1pt}
\begin{tabular}{l ccc c ccc c ccc c}
\toprule
\multirow{2}{*}{Model}
& \multicolumn{3}{c}{Endpoint}
& \phantom{a}
& \multicolumn{3}{c}{\texttt{Endpoint + keys}}
& \phantom{a}
& \multicolumn{3}{c}{\texttt{Full URL}} \\
\cmidrule(lr){2-4} \cmidrule(lr){6-8} \cmidrule(lr){10-12}
& P & R & F1 & & P & R & F1 ($\Delta$) & & P & R & F1 ($\Delta$) \\
\midrule
Qwen3-8B       & 33.6 & 27.9 & 30.5 && 31.2 & 26.1 & 28.4 ($-2.1$) && 23.7 & 23.5 & 23.6 ($-6.9$) \\
Ministral3-8B  & 23.2 & 33.5 & 27.4 && 19.3 & 28.7 & 23.1 ($-4.4$) && 13.1 & 25.7 & 17.3 ($-10.1$) \\
Gemma4-8B      & 45.2 & 12.2 & 19.2 && 43.3 & 11.9 & 18.7 ($-0.5$) && 39.5 & 10.7 & 16.9 ($-2.3$) \\
Qwen3-C80B$^{*}$ & 54.2 & 38.8 & 45.2 && 50.4 & 36.0 & 42.0 ($-3.2$) && 44.3 & 32.4 & 37.4 ($-7.8$) \\
Qwen3.6-35B & 51.6 & 39.2 & 44.6 && 50.5 & 38.3 & 43.5 ($-1.0$) && 47.9 & 34.2 & 39.9 ($-4.7$) \\
Llama4-S109B   & 40.0 & 34.8 & 37.2 && 35.6 & 32.3 & 33.9 ($-3.3$) && 31.2 & 28.8 & 30.0 ($-7.2$) \\
Qwen3-C480B$^{*}$ & 54.7 & 48.9 & 51.6 && 51.0 & 45.7 & 48.2 ($-3.4$) && 49.1 & 41.2 & 44.8 ($-6.8$) \\
\midrule
Gemini-Pro     & 66.3 & 63.8 & 65.1 && 63.9 & 61.7 & 62.8 ($-2.2$) && 60.2 & 60.0 & 60.1 ($-4.9$) \\
GPT-Terra & 70.5 & 60.3 & 65.0 && 68.8 & 58.7 & 63.3 ($-1.6$) && 66.2 & 53.9 & 59.4 ($-5.6$) \\
Gemini-Flash   & 69.7 & 53.8 & 60.7 && 65.4 & 50.6 & 57.1 ($-3.6$) && 54.7 & 47.9 & 51.1 ($-9.7$) \\
\bottomrule
\end{tabular}
\label{tab:url_strictness}
\end{table}

\section{False-positive Taxonomy: Labelling Rules}
\label{apx:fp_taxonomy}

We attribute each false positive (FP) to a single provenance category using a deterministic rule set.

\noindent\textbf{Notation.}
For sample $s$ and dimension $d \in \{\text{url},\text{domain},\text{ip},\text{file}\}$, let $G_{s,d}$ and $\hat{Y}_{s,d}$ be the canonicalized reference and prediction sets (canonicalization follows the headline scorer), and $F_{s,d} = \hat{Y}_{s,d} \setminus G_{s,d}$. Let $r$, $\ell$, and $\mathrm{ed}$ denote the \texttt{difflib} sequence-matcher ratio, the longest common contiguous substring length, and the Levenshtein edit distance.

\noindent\textbf{Closest reference.}
Each FP $f$ is paired with
\begin{equation}
g^{\star}(f) \;=\; \arg\max_{g \in G_{s,d}}\;
\max\!\Bigl(r(f,g),\; \ell(f,g)/\min(|f|,|g|)\Bigr).
\end{equation}

\noindent\textbf{Precedence.}
Predicates are evaluated top to bottom; the first match assigns the label, so every FP receives exactly one. Rules 1--2 handle malformed values, the rest well-formed ones. Category definitions and examples are in Table~\ref{tab:fp_taxonomy_v7}; each rule below states only its deciding predicate.
\begin{enumerate}\setlength{\itemsep}{1pt}
\item[1.] \textsc{Normalization error}: a bounded \emph{deterministic} normalization (scheme repair, leading-zero octet, case-fold) maps the ill-formed $f$ to a reference IOC.
\item[2.] \textsc{Ungrounded invalid}: $f$ is ill-formed and traces to neither gold nor source.
\item[3.] \textsc{Benign shared service}: $\mathrm{host}(f)$ is a known benign service and $f$ relates to \emph{no} reference IOC; source-absent benign services fall to rule 9 instead.
\item[4.] \textsc{Type mismatch}: normalized $f$ equals a reference value of a \emph{different} IOC type (an IP in the \texttt{domain} field); domain\,$\leftrightarrow$\,pathless-URL coincidences count as \textsc{Granularity}.
\item[5.] \textsc{Granularity}: the right entity at a different altitude, per type: the host or a path-prefix of a reference URL, or a parent/child of a reference domain; IPs admit exact matches only.
\item[6.] \textsc{Near-copy}: a syntactically valid distortion of $g^{\star}$: a same-type substring or superstring (a boundary mis-cut) or a fuzzy near-copy differing by a few characters ($r(f,g^{\star})\ge 0.90$ and $\mathrm{ed}\le 3$, or $r\ge 0.80$ with a long common substring). A cut \emph{inside} a DNS label is Near-copy; a cut at a path segment is Granularity.
\item[7.] \textsc{Prompt echo}: $f$ (or its host) occurs in the instruction template.
\item[8.] \textsc{Source non-IOC}: $f$ is grounded in the source (raw or decoded) but is neither an indicator nor a benign service.
\item[9.] \textsc{Generic placeholder}: a source-absent reserved/private/example pattern, or a benign service absent from the sample.
\item[10.] \textsc{Unattributed}: none of the above---a specific, valid-looking value with no trace in gold, source, template, or patterns.
\end{enumerate}
The \textbf{Disposition} column of Table~\ref{tab:fp_taxonomy_v7} gives each category's likely operational handling if the error survives triage; it is a rough severity ordering rather than a single-property axis (\textsc{Unattributed} spans O2--O3, since a plausible fabricated value both costs review time and can misdirect).

\begin{figure*}[t]
    \centering
    \includegraphics[width=0.75\linewidth]{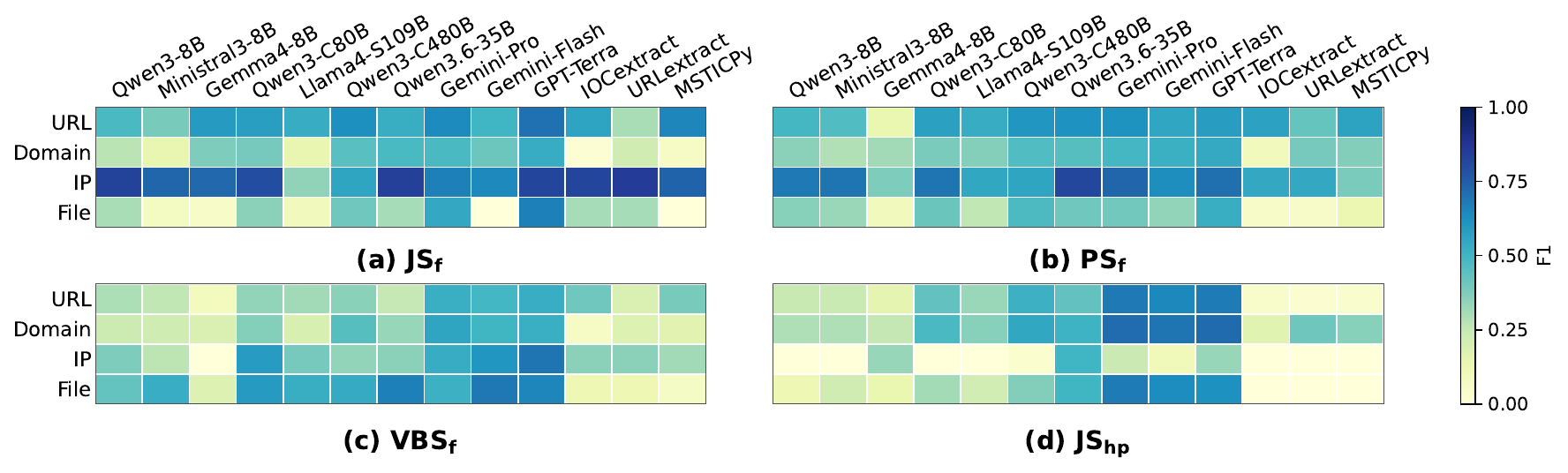}
    \caption{F1 score across models, datasets, and IOC types. Each cell shows the model performance for a specific IOC type. Darker colors indicate higher F1.}
    \label{fig:ioc_heatmap}
\end{figure*}

\noindent\textbf{Threshold sensitivity.}
The numerical cut-offs in the \textsc{Near-copy} rule (ratios of $0.90$ and $0.80$, edit distance $\le 3$, LCS length $\ge 8$, normalised LCS $\ge 0.6$) were tuned on a held-out development split of false positives drawn from a non-evaluation model (Qwen2.5-1.5B-Instruct). Shifting either ratio threshold by $\pm 0.05$ moves at most $2\%$ of reported FPs between adjacent IOC-related categories and never across the non-IOC / IOC boundary; the cut-offs are reported for reproducibility rather than as theoretically motivated constants.


\section{IOC-type-wise analysis}
\label{apx:ioc_analysis}
Figure~\ref{fig:ioc_heatmap} shows substantial variation across script and IOC types. URL and domain extraction is relatively stable for the frontier models, suggesting greater robustness to obfuscation and contextual ambiguity. Open-weight models vary more substantially, with notable drops on $JS_{hp}$, suggesting that their performance depends more strongly on indicators remaining close to surface form.

Filesystem artifacts are the most challenging IOC category overall, with consistently low F1 across models and datasets. Unlike URLs or IP addresses, file indicators are less structurally constrained: filenames, paths, extensions, temporary artifacts, payload names, and benign-looking strings may all coexist in the same script. Correct recovery therefore requires contextual interpretation of which strings correspond to actual malicious file activity, rather than surface-pattern matching alone.

Much of this difficulty is concentrated in the filename. A file indicator consists of a directory and basename, and either component may be assembled from split or transformed strings. Directories often follow a small set of familiar conventions (e.g., \texttt{\%TEMP\%}, \texttt{\%APPDATA\%}), whereas basenames are more often sample-specific and constructed through concatenation. Consistent with this distinction, directory-level scoring yields substantially higher F1 than full-path scoring: for GPT-Terra, 81.9 versus 61.4, and for Qwen3.6-35B, 67.8 versus 49.8. This consistent gap indicates that reconstructing the basename is the primary bottleneck in full-path recovery.

\section{IOC Recovery-Level Labeling}
\label{apx:recovery-levels}
Each ground-truth IOC is graded by \emph{how} it must be recovered from the script, using a fixed decoder as a reference model of \emph{mechanical} recovery. The decoder serves as a released, reproducible instrument against which the recovery gap achieved by LLMs can be measured (\S\ref{sec:eval}).

\noindent\textbf{Transform set.}
The decoder covers reversible, program-independent transformations commonly documented in prior work on malicious-script obfuscation and deobfuscation~\cite{bohannon2017revoke,ugarte2019powerdrive,li2019powershell,curtsinger2011zozzle,herrera2020safedeobs,xu2012power,barrsmith2021survivalism,mitre_t1027,iocextract}. As summarized in Table~\ref{tab:transform-battery}, \emph{Reassembly} operators reconstruct strings from characters already present in the source, while \emph{Transcode} operators invert reversible encodings or byte transformations. No operator interprets sample-specific control flow or dataflow.

We exclude program-dependent computation (loops, arithmetic, and control-flow-driven decoders) because recovering those requires executing attacker-controlled logic rather than inverting a fixed codec; that exclusion is what the \ltwo{} boundary encodes.

\begin{table}[t]
\centering
\footnotesize
\renewcommand{\arraystretch}{1.2}
\setlength{\tabcolsep}{4pt}
\begin{tabular}{@{}l p{0.62\linewidth}@{}}
\toprule
\textbf{Operator} & \textbf{What it does} \\
\midrule
\multicolumn{2}{@{}l}{\emph{Reassembly}: reorders characters already in the source} \\
\midrule
Literal concat & Joins adjacent string literals \\
Const.\ propagation & Single-assignment substitution of \texttt{var}$\,\leftarrow\,$literal or prior-variable concat; no control flow, loops, or computed values \\
Separator strip & Removes the most frequent repeated $n$-gram, $n\in[3,16]$ \\
Split/join & Rewrites \texttt{split}/\texttt{join} with literal arguments \\
NUL strip & Removes embedded NUL bytes \\
De-interleave & Unweaves interleaved character streams \\
Reversal & Reverses the character order of a string \\
\midrule
\multicolumn{2}{@{}l}{\emph{Transcode}: inverts a codec} \\
\midrule
Base64 & Standard alphabet and a 64-char alphabet declared in the source \\
Hex & Hexadecimal decoding \\
Char-code array & Decimal char codes: plain, offset, and sign variants \\
Percent & Percent-decoding \\
JS escapes & \texttt{\textbackslash NNN}, \texttt{\textbackslash uNNNN} \\
Base64\,$\rightarrow$\,UTF-16LE & Base64 decode reinterpreted as UTF-16LE \\
Single-byte XOR & Brute-forced over Base64/hex-decoded blobs under $4$\,KB \\
ROT-$n$ & $n\in\{13,47\}$ \\
Replace-chains & Literal replace-chains, $\le 40$ pairs \\
\bottomrule
\end{tabular}
\caption{Transform set applied by the fixed decoder. Operators are applied recursively to depth $4$ with hash-based de-duplication, a $4$\,MB per-blob cap, and $\le 60$ matches per operator, yielding candidate decoded blobs per script.}
\label{tab:transform-battery}
\end{table}

\noindent\textbf{Recovery levels.}
For each eligible (statically obtainable) IOC, we assign one of three levels:
\begin{center}
\footnotesize
\renewcommand{\arraystretch}{1.2}
\begin{tabular}{@{}l p{0.66\linewidth}@{}}
\toprule
\textbf{Level} & \textbf{Assigned when} \\
\midrule
\lzero{}  & The value appears verbatim in the source \\
\lone{}   & The value appears in some blob produced by the decoder \\
\ltwo{}   & Otherwise: recovery requires reasoning over the script's own decode logic \\
\bottomrule
\end{tabular}
\end{center}

For composite indicators, we assign the highest recovery level among their components. The labeling is deterministic and model-independent. Importantly, \ltwo{} is a one-sided label: it denotes values not reached by the fixed decoder, not values that are impossible to recover statically.

\subsection{Where Frontier Models Succeed and Fail}
\label{apx:transform-recall}
We further profile GPT-Terra, Gemini-Pro, and Gemini-Flash on the two harder recovery levels.

\smallskip
\noindent\textbf{\lone{} mechanisms.}
Table~\ref{tab:transform_recall} shows that frontier models are generally strongest on common structural reconstruction, such as split/join and literal concatenation, while decoding-heavy transformations such as Base64 are less reliable. Performance also varies substantially by operator and model.


\smallskip
\noindent\textbf{\ltwo{} mechanisms.}
Table~\ref{tab:l2_mechanism_recall} groups \ltwo{} IOCs by their dominant construction mechanism, including multi-variable dataflow, execution-mediated assembly, loop construction, nested encoding or compression, and arithmetic- or lookup-based construction. Unlike \lone{}, these indicators cannot be recovered through a predefined sequence of local transforms; recovery requires following computation specific to the sample.


\begin{table}[t]
\centering
\caption{Per-transform recall (\%) for the three frontier models over \lone{} IOCs.}
\label{tab:transform_recall}
\footnotesize
\setlength{\tabcolsep}{2.5pt}
\begin{tabular}{l r r r r}
\toprule
Transform & $n$ & GPT-Terra & Gem.-Pro & Gem.-Flash \\
\midrule
\multicolumn{5}{@{}l}{\emph{Reassembly (single)}} \\
\quad Literal concat        & 116 & 91.4 & 93.1 & 91.4 \\
\quad Split/join            &  83 & 100.0 & 85.5 & 83.1 \\
\quad Constant propagation  &  24 & 54.2 & 79.2 & 91.7 \\
\quad NUL strip             & \phantom{0}7 & 100.0 & 71.4 & 85.7 \\
\quad Reversal              & \phantom{0}2 & 100.0 & 100.0 & 100.0 \\
\midrule
\multicolumn{5}{@{}l}{\emph{Transcode (single)}} \\
\quad JS escapes            &  67 & 85.1 & 85.1 & 70.1 \\
\quad Base64\,$\to$\,UTF-16LE & \phantom{0}7 & 85.7 & 57.1 & 42.9 \\
\quad Base64                & \phantom{0}7 & 71.4 & 71.4 & 71.4 \\
\quad Single-byte XOR       & \phantom{0}2 & \phantom{0}0.0 & \phantom{0}0.0 & \phantom{0}0.0 \\
\midrule
\multicolumn{5}{@{}l}{\emph{Stacked ($\ge 2$ transform operations)}} \\
\quad Reassembly $+$ transcode & 165 & 66.1 & 66.1 & 51.5 \\
\quad Reassembly $\ge 2$       & \phantom{0}6 & 50.0 & 66.7 & 33.3 \\
\quad Transcode $\ge 2$        & \phantom{0}6 & 16.7 & \phantom{0}0.0 & \phantom{0}0.0 \\
\bottomrule
\end{tabular}
\end{table}

\begin{table}[t]
\centering
\caption{\ltwo{} recall (\%) for the three frontier models over \ltwo{} indicators, grouped by dominant construction mechanism (source-pattern heuristics).}
\label{tab:l2_mechanism_recall}
\footnotesize
\setlength{\tabcolsep}{2pt}
\resizebox{\columnwidth}{!}{%
\begin{tabular}{l r r r r}
\toprule
Mechanism & $n$ & GPT-Terra & Gem.-Pro & Gem.-Flash \\
\midrule
Multi-variable dataflow       & 362 & 30.9 & \textbf{49.7} & 29.6 \\
Execution-mediated            & 522 & 38.5 & \textbf{49.8} & 38.9 \\
Loop-constructed              & 150 & \phantom{0}9.3 & \textbf{16.7} & \phantom{0}4.0 \\
Nested encoding / compression & 143 & \textbf{51.7} & 44.1 & 31.5 \\
Computed-arithmetic / lookup  &  53 & 66.0 & \textbf{73.6} & 37.7 \\
\midrule
Overall \ltwo{}               & 1230 & 35.4 & \textbf{46.1} & 31.0 \\
\bottomrule
\end{tabular}%
}
\end{table}

\smallskip
\noindent\textbf{Recovery-Level Micro Recall by Dataset.}
Table~\ref{tab:recall_by_level_per_dataset} shows recall by recovery level within each dataset for three frontier models. 
The \lzero{}\,$>$\,\lone{}\,$>$\,\ltwo{} ordering holds in every dataset individually, so the recovery-level difficulty gradient tracks recovery level rather than dataset, collection period, or language.

\begin{table}[t]
\centering
\footnotesize
\setlength{\tabcolsep}{4.3pt}
\renewcommand{\arraystretch}{1.05}
\caption{Micro recall (\%) by recovery level within each dataset. L0 = \lzero{}, L1 = \lone{}, L2 = \ltwo{}.}
\label{tab:recall_by_level_per_dataset}
\begin{tabular}{l ccc ccc ccc}
\toprule
& \multicolumn{3}{c}{GPT-Terra} & \multicolumn{3}{c}{Gem.-Pro} & \multicolumn{3}{c}{Gem.-Flash} \\
\cmidrule(lr){2-4} \cmidrule(lr){5-7} \cmidrule(lr){8-10}
Dataset & L0 & L1 & L2 & L0 & L1 & L2 & L0 & L1 & L2 \\
\midrule
$JS_f$    & 95.4 & 33.3 & 10.7 & 89.8 & 44.4 & 13.1 & 51.9 & 22.2 & 11.9 \\
$PS_f$    & 75.6 & 58.8 & 20.0 & 77.5 & 50.0 & 10.0 & 75.0 & 47.1 & \phantom{0}6.0 \\
$VBS_f$   & 90.1 & 47.8 & 31.4 & 86.1 & 54.3 & 39.7 & 87.1 & 52.2 & \phantom{0}6.6 \\
$JS_{hp}$ & 96.5 & 86.1 & 38.9 & 93.9 & 83.9 & 51.6 & 96.1 & 75.7 & 36.9 \\
\bottomrule
\end{tabular}
\end{table}

\subsection{Case Study: Qwen3-8B Variants}
\label{apx:qwen-case-study}
\begin{figure}[t]
    \centering
    \includegraphics[width=\linewidth]{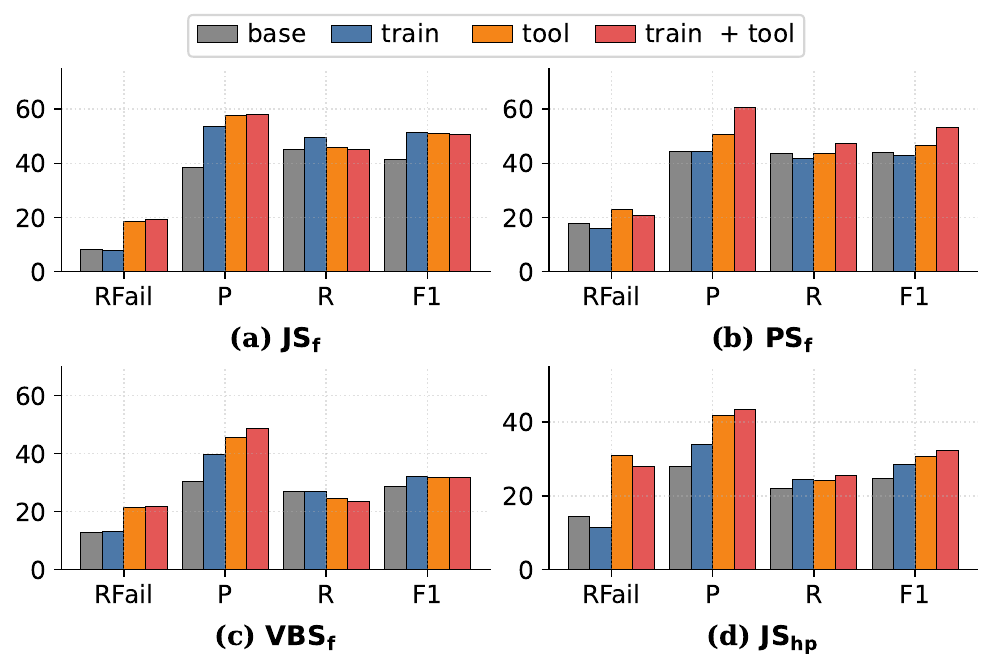}
    \caption{Effect of string utilities and task-specific adaptation on IOC extraction (Qwen3-8B variants), by script categories.}
    \label{fig:qwen3_variants}
\end{figure}

\end{document}